\documentclass[largeformat, a4paper]{interact}

\usepackage[numbers,sort&compress]{natbib}% Citation support using natbib.sty
\bibpunct[, ]{[}{]}{,}{n}{,}{,}% Citation support using natbib.sty
\makeatletter% @ becomes a letter
\def\NAT@def@citea{\def\@citea{\NAT@separator}}% Suppress spaces between citations using natbib.sty
\makeatother% @ becomes a symbol again

\theoremstyle{plain}% Theorem-like structures provided by amsthm.sty

\theoremstyle{definition}

\theoremstyle{remark}

\begin{document}

	\title{Laser cooling of molecules}
	
	\author{
		\name{M.~R. Tarbutt\thanks{Email: m.tarbutt@imperial.ac.uk}}
		\affil{Centre for Cold Matter, Blackett Laboratory, Imperial College London, London SW7 2AZ, UK}
	}
	
	\maketitle
	
	\begin{abstract}
	Recently, laser cooling methods have been extended from atoms to molecules. The complex rotational and vibrational energy level structure of molecules makes laser cooling difficult, but these difficulties have been overcome and molecules have now been cooled to a few microkelvin and trapped for several seconds. This opens many possibilities for applications in quantum science and technology, controlled chemistry, and tests of fundamental physics. This article explains how molecules can be decelerated, cooled and trapped using laser light, reviews the progress made in recent years, and outlines some future applications.
	\end{abstract}
	

\section{Introduction and motivation}\label{Sec:Intro}

Molecules in a room temperature gas move around at several hundred metres per second. It is difficult to make precise measurements with these molecules, because the high speed limits the observation time, and the wide range of Doppler shifts broadens their spectral lines. Molecules at room temperature also occupy a wide range of rotational and vibrational states, which smears out their quantum-mechanical properties. 

About 40 years ago, researchers began to find ways to cool {\it atoms} to very low temperatures using laser light~\cite{Phillips1998}. This early work on laser cooling was motivated by a desire to improve atomic clocks and spectroscopic measurements, but it was soon realised that the applications extended very much further. Ultracold atoms have been at the forefront of physics research ever since. They are used to test fundamental physics, determine fundamental constants, investigate ultracold collisions, study quantum degenerate gases and quantum phase transitions, simulate many-body quantum systems, store and process quantum information, and to measure time, gravity, acceleration, magnetic fields, electric fields, and pressure. 

For the first 20 years of this great flourishing, few gave much thought to the idea of cooling molecules. Physicists were having plenty of fun with atoms, and perhaps thought of molecules as unhelpful complexity. Chemists, though much more comfortable with molecules, were not easily convinced that it was interesting to study them at such low temperatures. Besides, applying laser cooling methods to molecules seemed, at first, prohibitively difficult. This situation has been transformed in recent years. Some groups developed methods to make ultracold molecules by binding together ultracold atoms~\cite{Moses2017}. This method has been tremendously successful for certain molecules, especially those formed from alkali atoms, with gases of polar molecules in the quantum degenerate regime now being formed this way~\cite{DeMarco2018}. Other groups showed how to decelerate molecular beams to low speed, and then confine those molecules for long periods in traps and storage rings~\cite{vandeMeerakker2012}. A few groups started to work out how to apply laser cooling methods directly to molecules, and that has now been done with spectacular success~\cite{Shuman2009,Shuman2010,Barry2012,Hummon2013,Zhelyazkova2014,Barry2014,McCarron2015,Yeo2015, Norrgard2016, Steinecker2016,  Hemmerling2016, Truppe2017, Truppe2017b,Williams2017,Anderegg2017,Kozyryev2017,Lim2018,Williams2018,McCarron2018, Anderegg2018, Cheuk2018, Collopy2018}. The applications of these ultracold molecules to a diverse range of topics is now within sight. The following paragraphs introduce some of these applications.

Molecules are already being used to test fundamental physics in a number of ways~\cite{DeMille2017,Safronova2018}. Several groups are using heavy polar molecules to measure the electron's electric dipole moment (edm). This is a fundamental property of the electron which is predicted by the Standard Model of particle physics to be exceedingly tiny. Popular extensions of the Standard Model, such as supersymmetry, predict much larger values, typically ten orders of magnitude larger. Although still tiny, these predicted values are within the reach of the most sensitive experiments. These experiments use molecules whose polar nature enhances the interaction energy between the edm and an applied electric field. Current measurements using warm molecules already place tight constraints on theories that extend the Standard Model~\cite{Hudson2011,Baron2014,Cairncross2017,Andreev2018}. Similarly, precise measurement with molecules can test whether the fundamental constants are changing with time or position~\cite{Shelkovnikov2008, Hudson2006, Truppe2013}. Such a change is predicted by some theories of dark energy, and theories that aim to unify gravity with the other forces. Molecules are also being used to study the influence of parity violation in nuclei~~\cite{Altuntas2018} and in chiral molecules~\cite{Tokunaga2013}. In all these experiments, the precision could be improved by using ultracold molecules to extend observation times and improve the degree of control~\cite{Tarbutt2009, Tarbutt2013, Cheng2016}.

Ultracold molecules can also bring advances in quantum science. Molecules can have large electric dipole moments that can be easily controlled, so they can interact quite strongly with one another through the long-range dipole-dipole interaction. Remarkable new phenomena, such as magnetism and superconductivity, emerge from collections of interacting quantum particles. These interacting many-body quantum systems are too complex to simulate, and are difficult to study in solids in a controlled way. As a result, the phenomena and the way they emerge, are not well understood. A regular array of polar molecules, with a single molecule on each site of the array and every molecule interacting with every other through the dipole-dipole interaction, makes an ideal system for learning about the physics of quantum many-body systems~\cite{Micheli2006,Barnett2006,Gorshkov2011,Yan2013,Blackmore2018}. To make the array, ultracold molecules can be loaded into an optical lattice, an array of traps made through the interference of overlapping laser beams. Quantum information processing is another area where molecules can have an impact, and several schemes have been proposed~\cite{DeMille2002,Yelin2006,Andre2006}. The rotational states can be the qubits, individual qubits can be controlled using microwave fields, and the dipole-dipole interaction can be used to build multi-qubit gates. The use of microwaves, rather than lasers, is possible for molecules because of the strong, electric dipole transitions between their rotational states, and is preferable because microwave fields are easier to control and to integrate within quantum circuits. For example, an interface that transfers quantum information between a molecule and a microwave photon seems feasible~\cite{Andre2006}.

Fundamental questions in chemical
physics can be addressed by cooling the reactants to low temperatures and preparing them in single quantum states~\cite{Krems2008, Richter2015, Liu2018}. In this regime, the wave nature of the reacting molecules, and the interference between these waves, is all important. Topics that could be studied include the influence of the vibrational, rotational and hyperfine state on the reactivity, the role of quantum tunnelling in reactions, and the influence of spin statistics on reactivity. Ultimately, it should be possible to control the outcome of a chemical reaction using static fields or lasers.

These diverse and exciting applications have motivated the community to develop techniques for producing the ultracold molecules needed. An important technique is direct laser cooling of molecules. This paper introduces the essential ideas of this topic, outlines the progress made in the field so far, and considers some likely new directions for the future. 

\section{Basics of laser cooling}
\begin{figure}[t]
	\centering
	\includegraphics[width=\columnwidth]{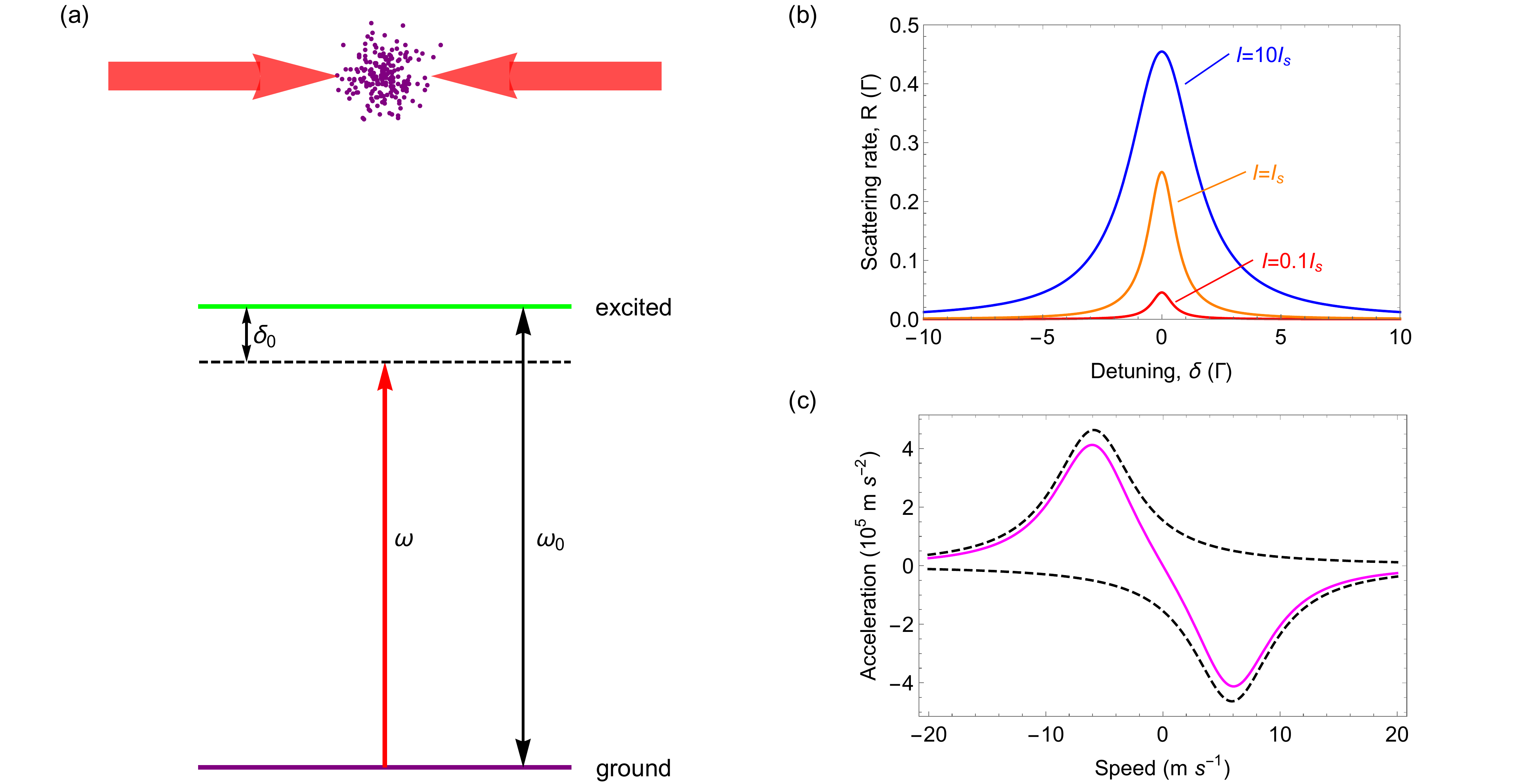}
	\caption{(a) Doppler cooling in 1D. Atoms interact with a pair of identical counter-propagating laser beams. The laser frequency is slightly below the atomic resonance frequency. (b) Scattering rate as a function of detuning of the laser light from resonance, for three different intensities. (c) Acceleration of a sodium atom as a function of its speed. Dashed lines show the accelerations due to each of the two beams, while the solid line is their sum. We have assumed the beams have intensities of $I=I_{\rm s}$, and are detuned by $\delta_0=-\Gamma$ from the $\lambda = 589$~nm transition (the yellow line of sodium).} \label{LaserCoolBasic}
\end{figure}

Let us begin by reviewing how Doppler cooling works for atoms. Figure \ref{LaserCoolBasic}(a) illustrates Doppler cooling of a simple atom in one dimension. The atom has a ground state and an excited state, and the angular frequency of the transition between these two states is $\omega_0$. The excited state has a lifetime $\tau$, and the transition has a corresponding natural linewidth $\Gamma = 1/\tau$. We note that $\Gamma$ is very small compared to $\omega_{0}$, typically $\Gamma \sim 10^{-8}\omega_{0}$ - atomic resonances are extremely sharp. The atoms interact with a pair of identical counter-propagating laser beams with wavevectors $\pm k$. The laser angular frequency, $\omega$, is tuned slightly below the atomic resonance angular frequency, so we say that the light is ``red-detuned''. The angular frequency difference is called the detuning, which we write as $\delta_0 = \omega - \omega_{0}$. The detuning is usually chosen to be of order $\Gamma$. Note that this simplified model where the atom has only two energy levels is usually sufficient. All the other levels of the real atom are irrelevant because the light is close to resonance with one particular transition, and very far from resonance with all the others.\footnote{An exception to this is the hyperfine structure, which is usually important in laser cooling. There may also be intermediate metastable states that have to be addressed.} 

For an atom moving with velocity ${\bf v}$, the two laser beams are not identical because there is a Doppler shift $\delta_{\rm D} = -{\bf k \cdot v}$, which is opposite for the two beams. A ground-state atom can be excited by absorbing a photon, most likely from the laser beam directed opposite to the motion, since this light is Doppler-shifted closer to resonance. The momentum of this absorbed photon is taken up by the atom, reducing its velocity by $\hbar k /m$, where $m$ is the atomic mass. The excited atom can decay back to the ground state by spontaneous emission. The recoil of the atom associated with this photon emission is in a random direction. This sequence - preferential absorption from the beam that opposes the motion followed by randomly-directed spontaneous emission - repeats until $v$ is close to zero.\footnote{When the laser intensity is uniform, sequences involving absorption followed by {\it stimulated} emission do not, on average, transfer any momentum to the atom, so we have not considered these processes in our discussion.} We refer to this process of absorption followed by spontaneous emission as photon scattering, and the associated force as the scattering force. 

For an atom interacting with a single laser beam of intensity $I$, the scattering rate is
\begin{equation}
R = \frac{\Gamma}{2}\frac{I/I_{\rm s}}{(1+I/I_{s}+4\delta^2/\Gamma^2)},
\label{eq:Rsc}
\end{equation}  
where $\delta = \delta_0 + \delta_{\rm D}$ is the total detuning, including the Doppler shift. The quantity $I_{\rm s}$ is known as the saturation intensity, and is defined as $I_{\rm s} = \pi h c \Gamma/(3 \lambda^3)$ where $\lambda$ is the wavelength. At small intensity, $I\ll I_{\rm s}$, $R$ increases linearly with $I$, while at large intensity, $I\gg I_{\rm s}$, $R$ saturates towards the maximum value $\Gamma/2$. Figure \ref{LaserCoolBasic}(b) plots Eq.(\ref{eq:Rsc}) as a function of $\delta$ for a few different values of $I/I_{\rm s}$. Naturally, the scattering rate is largest when $\delta = 0$, for then the light is resonant with the atom. The profile is a Lorentzian with a full width at half maximum of $\Gamma\sqrt{1+I/I_{\rm s}}$. At low intensity, this is simply the natural linewidth of the atomic transition, while at high intensity it is broadened because the scattering rate is more strongly saturated at the centre of the line than in the wings. From this dependence of $R$ on $\delta$, we can work out the scattering force, and how it depends on the speed of the atom. Force is the rate of change of momentum, and each scattering event transfers a momentum $\hbar k$, so the scattering force due to one beam is ${\bf F} = \hbar {\bf k} R$. When there are two counter-propagating beams, each of intensity $I$, the forces from each beam are opposite, but do not exactly cancel due to the opposite Doppler shifts. When $I \ll I_{\rm s}$, the total force is the sum of the two individual forces. Figure \ref{LaserCoolBasic}(c) shows the corresponding acceleration as a function of speed, in the case of sodium atoms. Here, we have chosen a laser detuning of $\delta_0 = -\Gamma$, and an intensity of $I=I_{\rm s}$ in each beam. The two dashed lines show the accelerations due to the two individual beams, and the solid line is the sum of these two.\footnote{Strictly, the total acceleration is only the sum of the individual accelerations when $I \ll I_{\rm s}$. Nevertheless, the plot is a good approximation for the higher intensities used here.} We see that the accelerations are very large, up to $4 \times 10^{5}$~m~s$^{-2}$. For small speeds, the acceleration varies linearly with speed and has the opposite sign to the speed, meaning it opposes the motion.  

At low speed, we can make a Taylor expansion of $F(v)$ around $v=0$, and thus obtain $F \approx -\alpha v$, where
\begin{equation}
\alpha = \frac{8 \hbar k^2 \delta_0 I/I_{\rm s}}{\Gamma(1+I/I_{\rm s}+4\delta_0^{2}/\Gamma^2)^2},
\label{Eq:alpha}
\end{equation}
is the damping constant. The atomic velocity is damped towards zero by this force. The acceleration is so large that the damping is very rapid - in our example of sodium, the characteristic damping time is just 14~$\mu$s. To extend the cooling into three dimensions, we introduce counter-propagating beams along each of the three orthogonal axes, so that all three components of the velocity are damped toward zero. This makes a three dimensional optical molasses.

In each scattering event, the spontaneous emission is in a random direction. When the speed is reduced almost to zero, absorption becomes equally likely from each of the beams, so the absorption events also have random directions. Thus, for each scattering event the atom receives two momentum kicks of $\hbar k$, each randomly directed. This random jostling of the atom is a heating mechanism which counteracts the damping towards zero velocity. There is an equilibrium temperature, known as the Doppler temperature, where the heating and cooling are balanced. The Doppler temperature is 
\begin{equation}
T_{\rm D} = - \frac{\hbar \Gamma^{2}}{8 k_{\rm B} \delta_0}\left(1+\frac{I}{I_{\rm s}}+\frac{4\delta_0^2}{\Gamma^2}\right).
\label{Eq:DopplerTemperature}
\end{equation}
When $I\ll I_{\rm s}$ and $\delta_0 = -\Gamma/2$, we get the minimum temperature, $T_{\rm D,min} = \hbar \Gamma/(2 k_{\rm B})$, known as the Doppler limit. For sodium, its value is 240~$\mu$K. As will be explained in section \ref{Sec:DoppSubDopp}, there are other cooling mechanisms that can reach temperatures below this limit, and atoms are routinely laser cooled to lower temperatures using these sub-Doppler techniques.

\section{Applying laser cooling to molecules}\label{Sec:LaserCoolingMolecules}

The change in velocity upon absorption of a photon, known as the recoil velocity, is $v_{\rm r} = \hbar k/m$. For sodium, the value is $v_{\rm r}=0.03$~m/s, and it is similar for other atoms and molecules. A typical speed for atoms or molecules in a room temperature vapour is $\sim 300$~m/s, so it takes at least $10^{4}$ scattering events to bring this speed close to zero. This requires a ``closed'' transition, meaning that the excited atom cannot leave the cooling cycle by decaying to some intermediate state that is not resonant with the laser light. Some atoms have these closed transitions, so are easy to cool. Molecules have the added complication of vibrational and rotational energy levels. To understand how to apply laser cooling to molecules, we need to examine this energy level structure.

\subsection{Energy level structure}

\begin{figure}[t]
	\centering
	\includegraphics[width=\columnwidth]{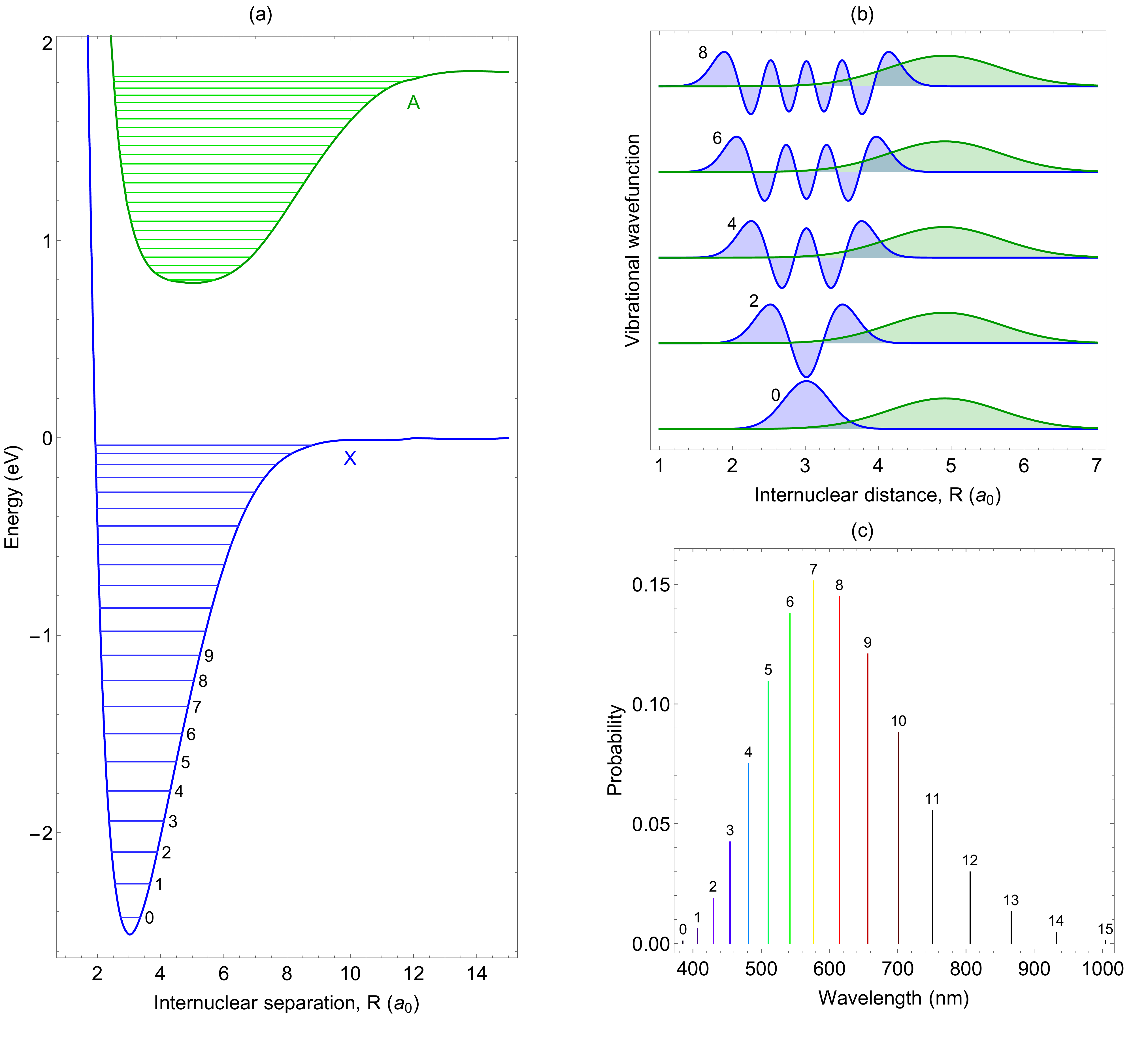}
	\caption{(a) Potential energy curves of the ground electronic state ($X$) and first electronically-excited state ($A$) of a typical diatomic molecule - the example here is LiH. The vibrational energy levels within each potential are shown. (b) Lowest vibrational wavefunction ($v'=0$) for the $A$ state (green), and a selection ($v''=0,2,4,6,8$) of vibrational wavefunctions for the X state (blue). The square of the overlap integral between a vibrational wavefunction from the $X$ state and one from the A state gives the corresponding Franck-Condon factor. (c) Emission spectrum for molecules excited to the $v'=0$ vibrational level of the A state.} \label{LiH_example}
\end{figure}

Figure \ref{LiH_example}(a) illustrates the energy level structure of a typical diatomic molecule. Molecules have a set of electronic states, just like atoms. The ground electronic state is always labelled $X$, while the excited electronic states are labelled alphabetically, $A$, $B$, $C$ etc, ordered by energy. For each electronic state, the electronic energy, $E_e$, is a function of the distance between the nuclei, $R$, so each electronic state is plotted as a curve $E_e(R)$. This is known as the potential energy curve, because it acts as the potential in which the nuclei move. The nuclei oscillate about the equilibrium separation, $R_0$, leading to a set of quantised vibrational energy levels, labelled by the quantum number $v=0,1,2...$.  These vibrational energy eigenvalues, $E_v$, and the corresponding eigenfunctions, $\Phi_{e,v}(R)$, are found by solving the one-dimensional Schr\"odinger equation for the potential energy curve $E_e(R)$. Near the bottom of the potential energy curve the motion is close to that of a harmonic oscillator, so the vibrational energy levels are $E_v \approx (v+1/2)\hbar\omega_{\rm vib}$, and the vibrational wavefunctions are the corresponding harmonic oscillator eigenfunctions. Figure \ref{LiH_example}(b) shows some example vibrational wavefunctions, which are discussed in more detail below. For each vibrational state, there is a set of rotational states. Recall from classical mechanics that the energy of a rotating body is $L^{2}/2I$, where $L$ is the angular momentum and $I$ is the moment of inertia of the body. Quantizing the angular momentum replaces the continuous variable $L^2$ with $\hbar^{2}N(N+1)$, where $N=0,1,2...$ is the rotational quantum number, so the rotational energies are $E_r=B N(N+1)$ where $B=\hbar^2/2I$ is known as the rotational constant of the molecule. The spacing of rotational levels is typically much smaller than the spacing of vibrational levels, which in turn is much smaller than the spacing of the potential energy curves corresponding to different electronic states. We can summarise this hierarchy as $\Delta E_r \ll \Delta E_v \ll \Delta E_e$.

\subsection{Vibrational branches}

Knowing the energy level structure, we can now consider how to apply laser cooling to molecules. We imagine driving a transition from the electronic ground state, $X$, with vibrational and rotational quantum numbers $v$ and $N$, labelled $|g\rangle=|X, v, N\rangle$, to a specific vibrational ($v'$) and rotational ($N'$) level of the first excited electronic state, $A$, labelled $|e\rangle=|A, v', N'\rangle$. For laser cooling to be efficient, we would like $|e\rangle$ to decay exclusively to $|g\rangle$ so that the molecule remains resonant with the laser light, allowing it to scatter a large number of photons from the laser. Unfortunately, that is not normally what happens. Instead, $|e\rangle$ can decay to any of the many vibrational states of $X$. The probability for decaying to $v''$ is\footnote{This is an approximation, which is usually sufficiently accurate when the wavefunctions are well localised around the equilibrium bond length.} 
\begin{equation}
P_{v',v''} = \left|\int \Phi_{X,v''}(R) \Phi_{A,v'}(R) d R\right|^2.
\label{Eq:FC}
\end{equation} 
This is the square of the overlap integral between two vibrational wavefunctions, one from the X state and one from the A state, and is known as the Franck-Condon factor.

\begin{figure}[t]
	\centering
	\includegraphics[width=\columnwidth]{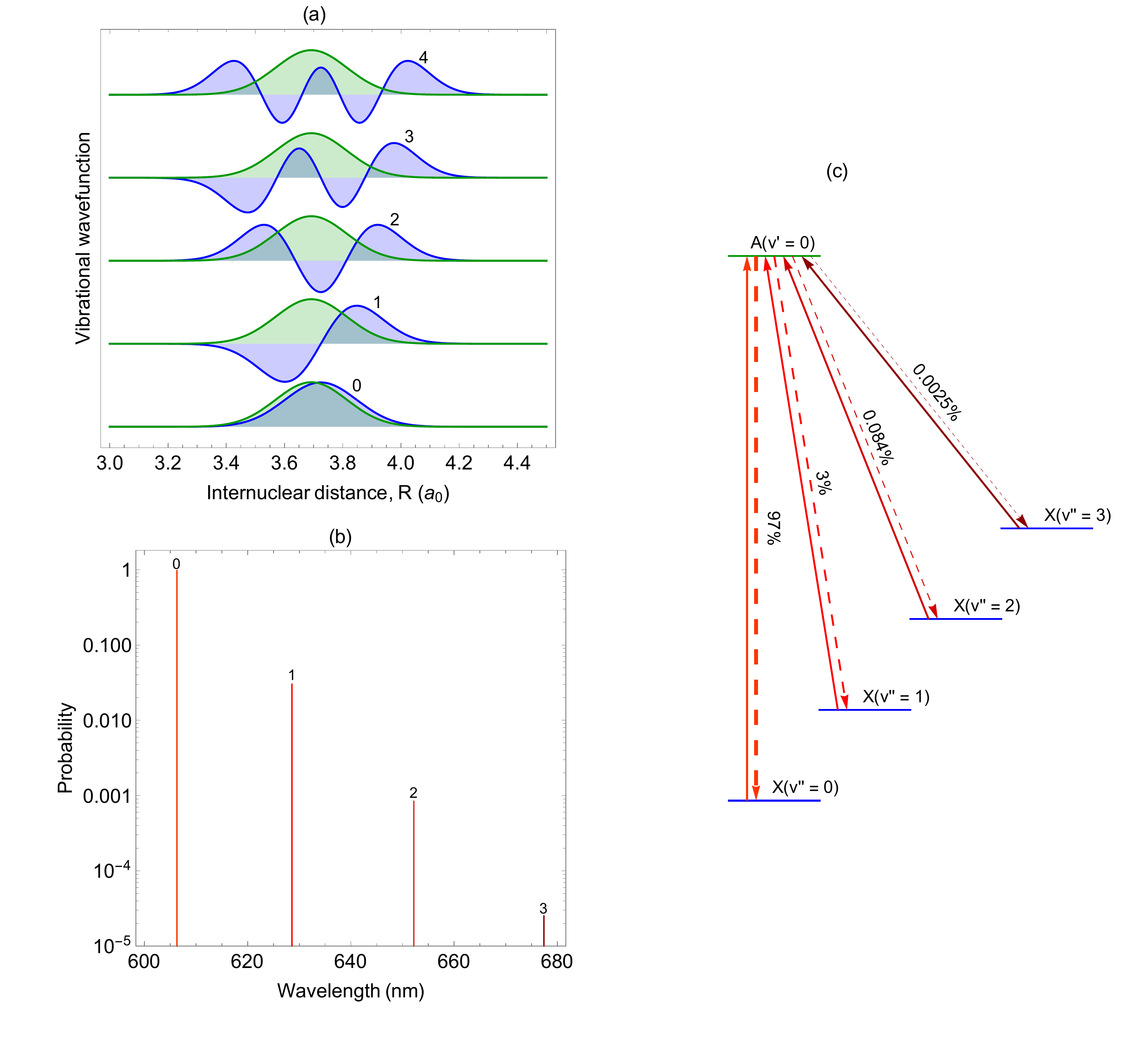}
	\caption{Some molecules, such as CaF, have vibrational branching ratios suitable for laser cooling. (a) Lowest vibrational wavefunction ($v'=0$) for the $A$ state of CaF (green), and a selection ($v''=0,1,2,3,4$) of vibrational wavefunctions for the $X$ state (blue). (b) Emission spectrum for molecules excited to the $v'=0$ vibrational level of the $A$ state. Note the logarithmic scale. (c) Laser cooling scheme for CaF involving four lasers.} \label{CaF_example}
\end{figure}

Figure \ref{LiH_example}(b) illustrates (in blue) the first five even-numbered\footnote{The even-numbered ones are chosen to illustrate the progression up to $v''=8$ without overly cluttering the figure; the odd-numbered ones are just as important!} vibrational wavefunctions, $\Phi_{X,v''}$, for the $X$ state potential shown in figure \ref{LiH_example}(a). For each one, we also show (in green) the lowest vibrational wavefunction of the $A$ state, $\Phi_{A,v'=0}$. A molecule excited to this state decays to the various vibrational states of $X$ with probabilities $P_{0,v''}$ given by equation (\ref{Eq:FC}). For $v''=0$, the overlap integral is given by the overlapping green and blue areas in the bottom row of figure \ref{LiH_example}(b). There is not much overlap because the equilibrium bond length of the two potential curves is quite different, so only the wings of the two ground-state wavefunctions overlap. As a result, $P_{0,0}$ is small. For $v''=2$, the integral is the overlapping green and blue areas in the next row up. There's still not much overlap, though it's a bit more than for $v''=0$. We see that, in this example, the excited molecule decays to many different vibrational states, with a fairly small probability to each. As a result, the excited molecule emits at many different wavelengths across the visible spectrum and beyond, as illustrated by Figure \ref{LiH_example}(c) which shows the emission spectrum for molecules excited to the $|A,v'=0\rangle$ state. Effective laser cooling requires about $10^{4}$ scattered photons, so all $v''$ with $P_{0,v''} > 10^{-4}$ have to be addressed. That means we need a laser at every one of the wavelengths shown in Figure  \ref{LiH_example}(c). This is impractical.

Fortunately, not all molecules are so obstructive. It is easy to see that if the X and A states have identical potential energy curves, $P_{v',v''}$ will be equal to 1 when $v''=v'$ and zero otherwise.\footnote{When the potential energy curves are identical, vibrational wavefunctions with $v''\ne v'$ are orthogonal.} In that case, a vibrational state of $A$ can decay to only one vibrational state of $X$. Some molecules come close to this ideal. Typically, they are ones where the electron that gets excited in the transition does not participate in the molecular bond. Then, the transition involves very little change to either the length or strength of the bond, and the potential energy curves for the ground and excited states will be very similar. Calcium monofluoride, CaF, is a good example of such a molecule. To a good approximation, one of the two valence electrons of calcium goes over to the fluorine to make the ionic bond, while the other remains localised on the calcium and is the electron that gets excited in the electronic transition between the $X$ and $A$ states. Figure \ref{CaF_example}(a) shows (in blue) the wavefunctions for the lowest 5 vibrational levels ($v''=0$ to 4) in the X state of CaF, and (in green) the ground vibrational wavefunction ($v'=0$) of the $A$ state. Because the potential energy curve of $X$ and $A$ are very similar, there is a large overlap between the $v'=0$ and $v''=0$ states. The overlap with all other states is very small\footnote{To see this, don't forget to include the sign of the wavefunction in the overlap integral!}. Figure \ref{Eq:FC}(b) shows the resulting emission spectrum for molecules excited to $|A,v'=0\rangle$. Note here the logarithmic scale on the vertical axis, spanning 5 orders of magnitude. We see that $P_{0,0} \approx 1$, and that $P_{0,0} \gg P_{0,1} \gg P_{0,2} \gg P_{0,3}$. This is one of the key ingredients for practical laser cooling of molecules~\cite{DiRosa2004}. In the example of CaF, shown in figure \ref{CaF_example}(c), four lasers are used to address the transitions from $v''=0,1,2,3$. The remaining leak out of the cooling cycle is then so small that molecules will, on average, scatter about $2 \times 10^{5}$ photons, sufficient for effective laser cooling.

A careful choice of molecule, and the use of several lasers to address enough vibrational transitions, solves the problem of decay to multiple vibrational states. Next, we turn to the rotational structure.

\begin{figure}[t]
	\centering
	\includegraphics[width=0.6\columnwidth]{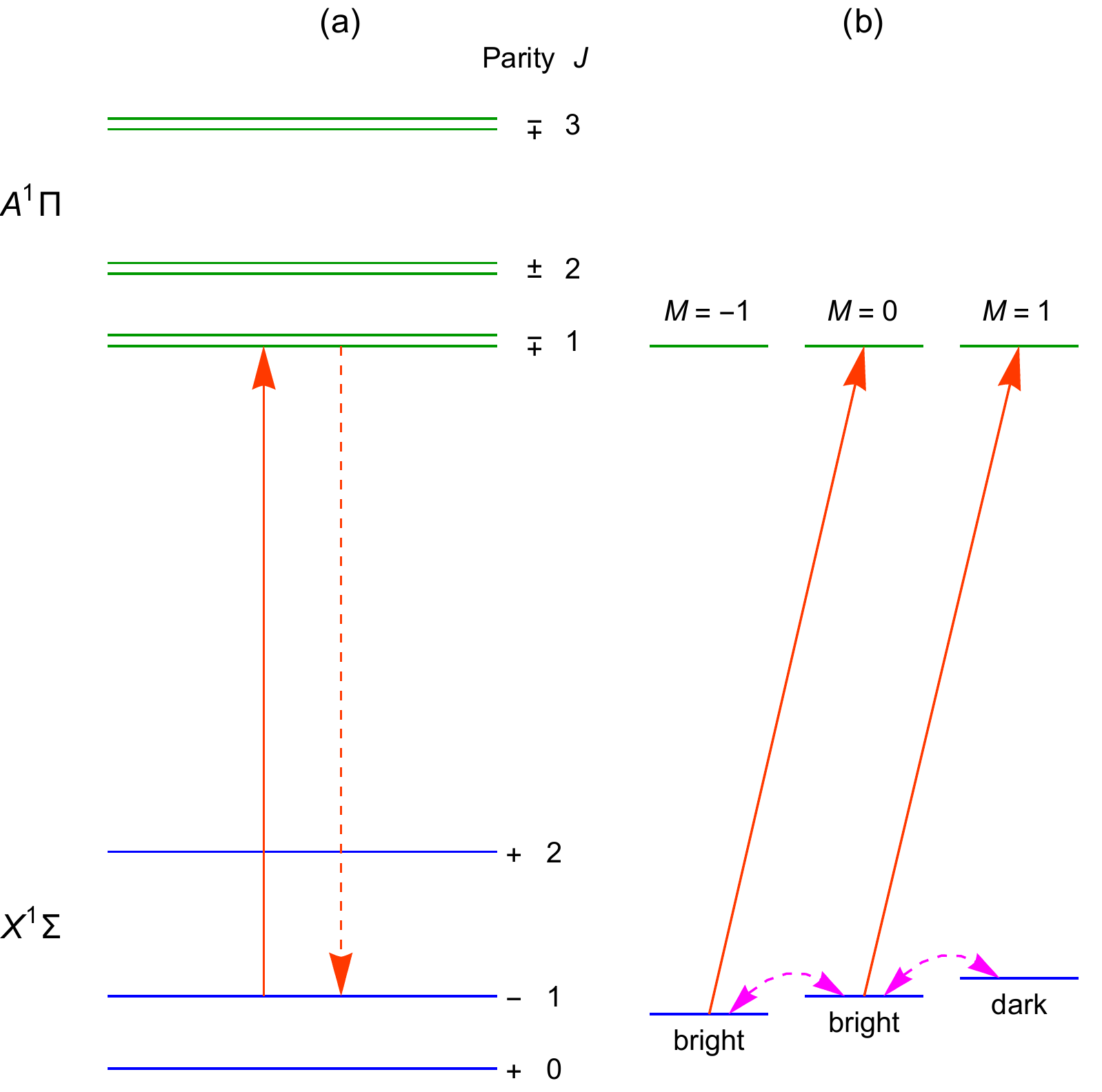}
	\caption{(a) Rotational structure of a typical laser cooling transition for a diatomic molecule (e.g. any one of the transitions illustrated in figure \ref{CaF_example}(c)). The labels specify the parity and total angular momentum ($J$) for each state. The transition indicated by the solid arrow is ``rotationally closed''. (b) Zeeman sub-levels of the $J=1$ lower and upper states. The component of angular momentum along the $z$-axis is specified by $M$. The arrows show the transitions that can be driven by left-handed circularly-polarised light propagating along the z-axis. For this choice of polarisation, the $M=1$ state is a dark state. A different polarisation choice will produce a different dark state. In a magnetic field, the Zeeman sub-levels have different energies and a dark state that is a superposition of these states is unstable (illustrated by magenta arrows).} \label{Rotational}
\end{figure}

\subsection{Rotational branches}

Within each vibrational state, there is a ladder of rotational states. We have to consider which of these rotational states might be populated when the molecule decays to the ground electronic state. Fortunately, the number of accessible rotational states is limited by the angular momentum and parity\footnote{The eigenfunctions, $\psi({\bf r})$, of a molecule have the property $\psi(-{\bf r})=p\times\psi{({\bf r})}$ where $p=\pm 1$ is the parity.} selection rules. Figure \ref{Rotational}(a) shows the typical rotational structure of the ground and excited states connected by the laser cooling transition. To keep things relatively simple, we consider a case where there is no unpaired electron spin and no unpaired nuclear spin. Then, the angular momentum comes only from the orbiting electrons and from the rotation of the molecule as a whole. Because of the molecule's symmetry, it is usual to refer to the projection of the electronic orbital angular momentum onto the internuclear axis. In our example, and in units of $\hbar$, this is 0 for the $X$ state and 1 for the $A$ state. These are known as $\Sigma$ and $\Pi$ states and are analogous to the S and P states of atoms. We add together the orbital and rotational angular momenta to give the total angular momentum quantum number $J$. In the $X$ state, the lowest level has $J=0$, and the parity alternates between positive and negative as we go up the rotational ladder. In the $A$ state, the lowest level has $J=1$, since this is the smallest value of $J$ that can give a projection of 1 onto the internuclear axis. In the $A$ state, there are a pair of states of opposite parity, $p$, for each value of $J$. This is because the projection of the orbital angular momentum onto the internuclear axis can either be $+1$ or $-1$, and the energy eigenstates are the symmetric and antisymmetric linear combinations of these possibilities.\footnote{These states with the same $J$ but opposite parity are not quite degenerate because of a slight, parity-dependent, mixing of $\Pi$ and $\Sigma$ states by the rotational motion (and the spin-orbit interaction for molecules with spin).}

Let us indicate angular momentum and parity using the notation $J^p$. The selection rules for electric dipole transitions require that the parity changes in the transition and that $\Delta J = 0,\pm 1$. Consider first the only allowed transition from the lowest level of $X$. The excited state of this transition is $1^-$. This state can decay back to $0^+$, but can also decay to $2^+$. The presence of these two decay channels is inconvenient for laser cooling, since it doubles the number of laser frequencies needed. Consider instead the transition from $1^-$ to $1^+$. Here, the upper state can only decay back to the $1^-$ level of $X$, so this transition is ``rotationally closed'', meaning that only one rotational component needs to be addressed (for each vibrational state)~\cite{Stuhl2008}. All laser cooling work to date has made use of such rotationally closed transitions.

\subsection{Dark states}\label{Sec:dark}

There is one last problem to solve. The laser cooling transition indicated in figure \ref{Rotational}(a) has $J=1$ in both the ground and the excited state. Each has 3 Zeeman sublevels, with quantum numbers $M = -1,0,+1$, as shown in figure \ref{Rotational}(b). The $M$ quantum number expresses the angular momentum of the atom in the $z$ direction. Suppose we have a circularly polarised laser beam propagating along the $z$-axis. The photons of this beam have one unit of angular momentum along $z$. When the atom is excited, by absorbing a photon, $M$ must increase by one in order to conserve the angular momentum along $z$, but this cannot happen for molecules with $M=+1$, since the excited state does not have an $M=+2$ state. Molecules in $M=+1$ cannot scatter any photons from the laser, and once they reach this state they remain there indefinitely. We call it a dark state. Molecules in other states can be excited by the laser, but the excited state can decay to $M=+1$, so every molecule will end up in the dark state after scattering just a few photons, preventing laser cooling. We have illustrated this problem for one choice of polarization, but it is actually very general. Had we chosen circular polarisation with the opposite handedness, the $M=-1$ state would be dark, and if we choose light linearly polarised along $z$, the $M=0$ state is dark. More generally, for any choice of laser polarisation, there will be a coherent superposition of the ground state sub-levels that is a dark state, and molecules will always reach this dark state after scattering just a few photons.

There are two main ways to solve this problem. One is to switch the polarization of the light between two orthogonal states, e.g. from left-circular to right-circular. The state which is dark to one polarization will not be dark to the other, so if the modulation is done rapidly enough, molecules will never be in a dark state for long. The relevant timescale is the time taken to reach the dark state, which in a typical experiment is about 1~$\mu$s. A second option is to choose the polarization so that the dark state is a superposition state, and apply a small magnetic field to `de-stabilise' that superposition. In a magnetic field, the energy depends on $M$, so the states that make up the superposition have different energies (see figure \ref{Rotational}(b)). According to the time-dependent Schr\"odinger equation, an initial dark superposition state $|1\rangle + |2\rangle$ evolves into the state $|1\rangle + e^{-i \Delta t/\hbar} |2\rangle$ after a time $t$, where $\Delta$ is the energy difference between states $|1\rangle$ and $|2\rangle$. Because of this change of phase, the new state is no longer dark, so molecules will not be in a dark state for long provided $\Delta$ is large enough. In most cases, a small magnetic field, just a few times the Earth's field, is enough to destabilise any dark state efficiently.

\subsection{Doppler and sub-Doppler cooling}\label{Sec:DoppSubDopp}

The recipe for laser cooling is now clear. Choose a molecule with a strong electronic transition that has favourable vibrational branching ratios, use enough vibrational repump lasers to reduce the leak out of the cooling cycle to an acceptable level (usually around $10^{-5}$), drive the rotationally closed transition indicated in figure \ref{Rotational}, and de-stabilise dark states by switching the polarization of the light or applying a magnetic field. The interaction of the molecule with the light is then very similar to that of the simple two-level system illustrated in figure \ref{LaserCoolBasic}. In particular, the formulae for the scattering rate, damping constant and Doppler temperature [equations (\ref{eq:Rsc}), (\ref{Eq:alpha}) and (\ref{Eq:DopplerTemperature})] are almost unchanged.\footnote{When multiple ground states are connected by lasers to a single excited state, as is often the case, the formulae are altered slightly. Most notably, the scattering rate is reduced. See, for example, reference \cite{Williams2017}.} For red-detuned light, we can expect Doppler cooling to cool the molecules close to the Doppler limit, just as it does for atoms.

\begin{figure}[t]
	\centering
	\includegraphics[width=0.5\columnwidth]{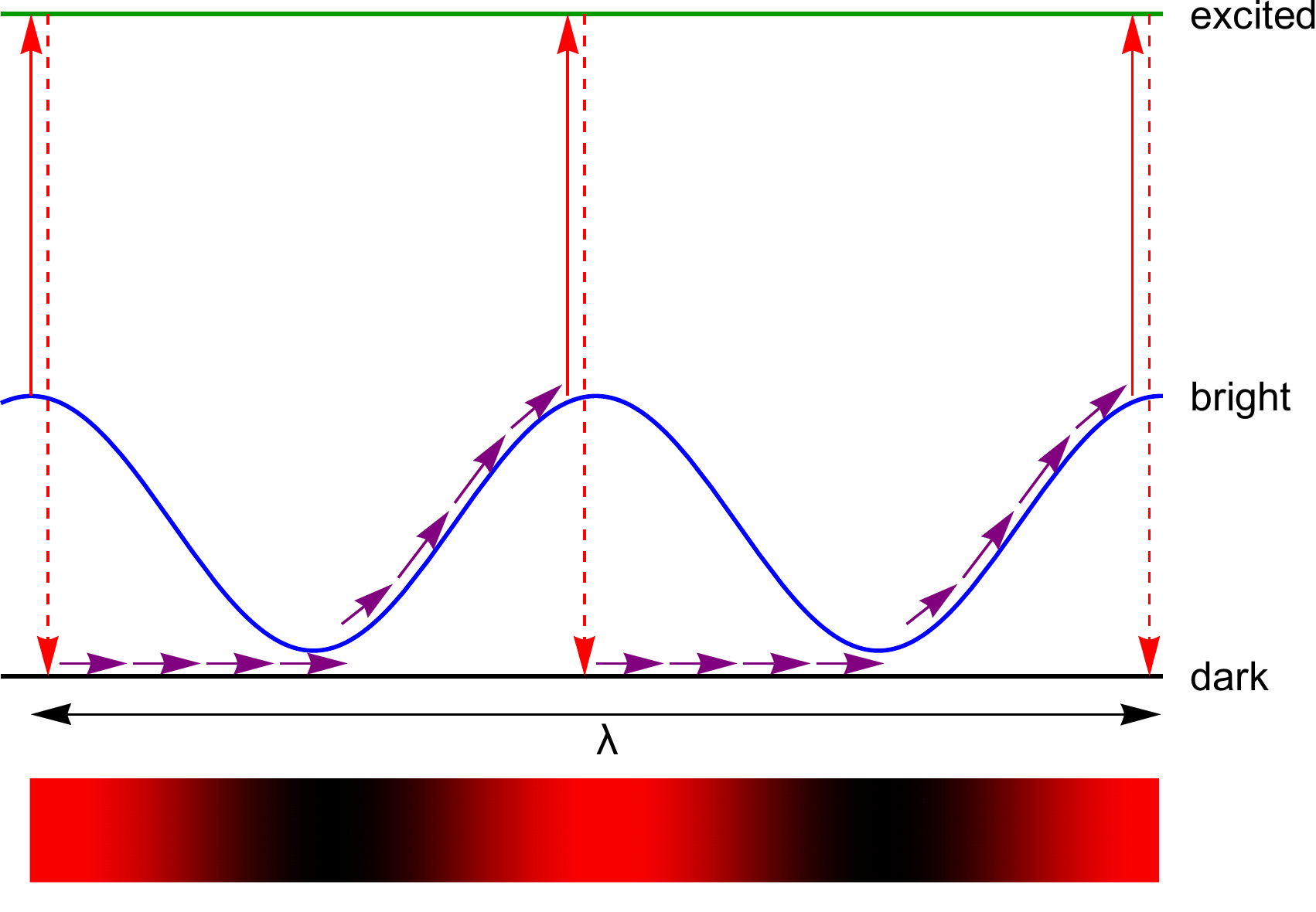}
	\caption{The Sisyphus cooling mechanism.} \label{SisyphusFigure}
\end{figure}

In practice, lower temperatures than the Doppler limit are routinely reached for atoms, because there are other cooling mechanisms, distinct from Doppler cooling, that become effective at low temperatures. These are known as sub-Doppler cooling mechanisms, and there are several different types~\cite{Dalibard1989,Ungar1989}. In the most effective of them, dark states are deliberately set up and exploited, showing that although dark states can be troublesome for laser cooling, they can also be put to good use~\cite{Weidemuller1994}. The mechanism is illustrated in figure \ref{SisyphusFigure}, where for simplicity we consider a molecule with a single excited state and two ground sub-levels. The laser light can excite a molecule from one of these sub-levels (the bright state), but not from the other (the dark state). Let us consider a pair of counter-propagating laser beams, slightly detuned from resonance. If the two beams have the same polarisation, they set-up an intensity standing wave with a period of $\lambda/2$, as illustrated at the bottom of the figure. If the polarisations are orthogonal, the intensity is uniform but the local polarisation changes with position, with a period of $\lambda$. If the polarisations of the two beams are neither parallel nor perpendicular, both the intensity and the local polarisation change with position. The energy levels of the molecule are influenced by the light. The oscillating electric field of the light drives the electrons in the molecule, creating an electric dipole oscillating at the same frequency. The interaction between this dipole and the electric field of the light results in a slight shift of the ground-state energy levels, negative when the light is red-detuned and positive when it is blue-detuned. This is known as the light shift or the ac Stark shift. The energy of the bright state is shifted by this effect, but the dark state does not shift because it does not interact with the light. In figure \ref{SisyphusFigure}, we have illustrated the case where the light is blue detuned, so the bright state lies above the dark state. In the standing wave, the shift of the bright state changes with position, being largest where the intensity is greatest\footnote{The shift may also depend on polarisation.}. Consider a molecule in the bright state, near a node of the standing wave. As it moves towards the antinode, its internal energy increases due to the ac Stark shift, so its kinetic energy must decrease. We think of the molecule slowing down as it climbs the potential hill. The molecule is most likely to be excited when it is near the antinode, for here the light is most intense, so it tends to be optically pumped to the dark state once it is near the top of the potential hill. The molecule can be transferred back to the bright state by two processes closely related to the two methods for de-stabilising dark states outlined in section \ref{Sec:dark}. If there is a magnetic field of an appropriate size, it will transform the dark state into the bright state. Alternatively, the motion of the molecule through the changing polarisation of the light can transfer the molecule back to the bright state. This transfer is most likely to happen where the bright and dark states are close in energy, which is near the bottom of the potential hill. As a result, the molecule tends to follow the purple arrows in the figure, losing energy as it climbs the potential hills over and over again. This is known as Sisyphus cooling. Furthermore, because the molecules spend much of their time in the dark state, the photon scattering rate is reduced, and so too is the heating due to the random jostling of photon scattering events. This, therefore, is a powerful mechanism for cooling to temperatures below the Doppler limit. 

\subsection{Progress in laser cooling of molecules}

The pioneering work on laser cooling of molecules was done by a team at Yale, using strontium monofluoride (SrF) molecules. For this molecule, the branching ratio for the main transition, $A(v'=0) \leftrightarrow X (v''=0)$ is 98\%. With a single vibrational repump laser addressing the leak to $v''=1$, the team showed that the molecules were scattering about 150 photons during a $50~\mu$s interaction time, enough to change their velocity by 0.8~m/s~\cite{Shuman2009}. 

\begin{figure}[t]
	\centering
	\includegraphics[width=\columnwidth]{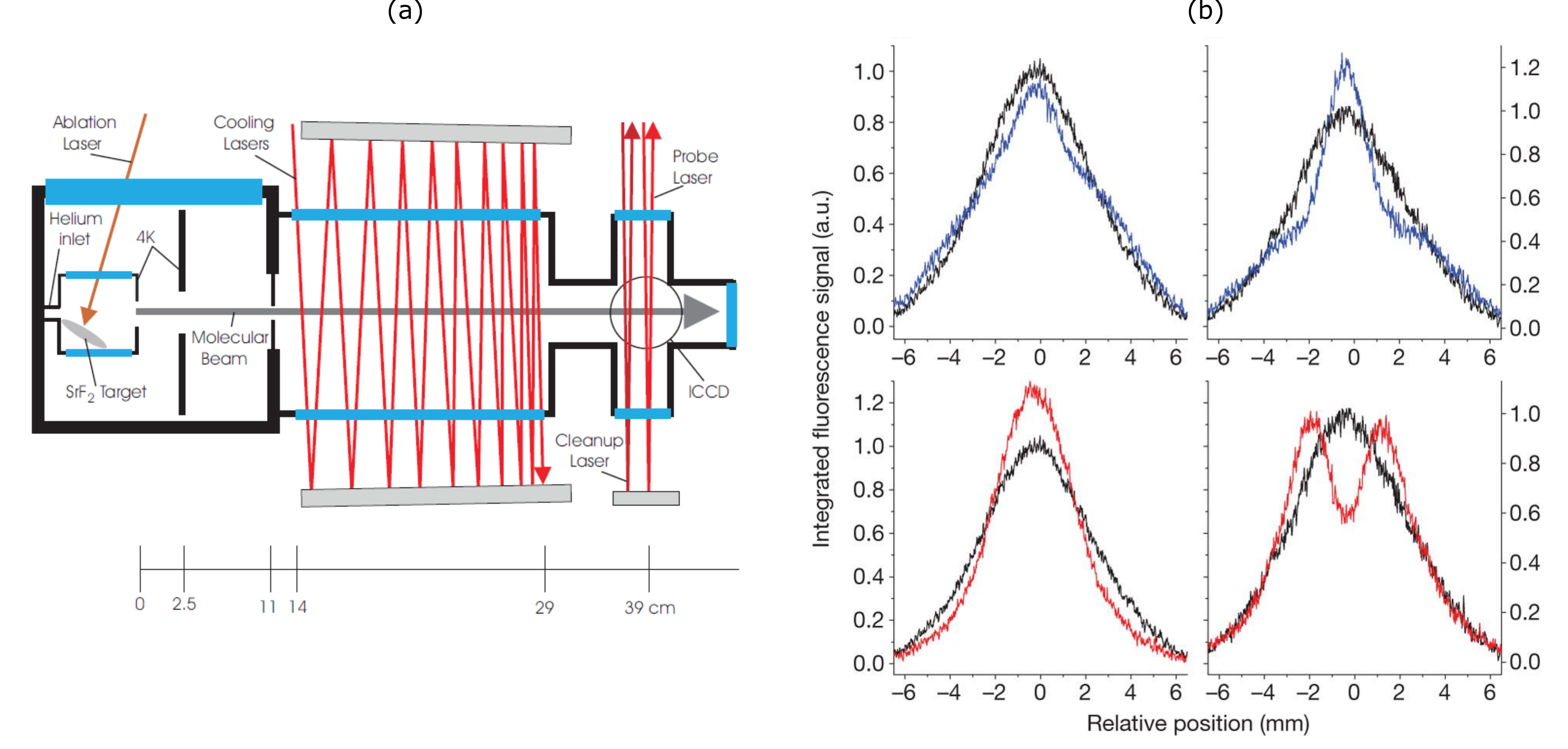}
	\caption{Transverse laser cooling of a SrF molecular beam. (a) Experimental setup. A molecular beam from a cryogenic buffer-gas source passes through a one-dimensional optical molasses formed by multiple passes of the laser light, and is then imaged onto a camera. (b) Resulting density distributions of the molecular beam. Black curves: no cooling light; red curves: detuning of $\delta_0=-1.5\Gamma$; blue curves: detuning of $\delta_0=+1.5\Gamma$. Left: with a magnetic field of 0.5~mT, Doppler cooling is observed for red-detuned light (lower graph) and Doppler heating for blue-detuned light (upper graph). Right: with a magnetic field of 0.06~mT, magnetically-assisted Sisyphus cooling is observed for blue-detuning (upper), and Sisyphus heating for red-detuning (lower). Adapted from reference \cite{Shuman2010} with permission from Springer Nature. } \label{SrFTransverse}
\end{figure}

Next, the Yale group added a second repump laser to close the leak to $v''=2$, so that far more photons could be scattered. Using this laser light, they formed a one-dimensional optical molasses and used it to cool a beam of SrF molecules in one transverse direction (one of the directions perpendicular to the propagation direction of the beam)~\cite{Shuman2010}. The experiment is illustrated in figure \ref{SrFTransverse}(a). A cryogenic buffer gas source was used to make a molecular beam, and the laser beams were reflected back and forth at a slight angle so that they intersected the molecular beam many times, forming a 15~cm long interaction region. After this cooling region, the density distribution of the molecular beam was measured. Figure \ref{SrFTransverse}(b) shows the results of these experiments for various cases. The graph on the bottom left shows the result for a red-detuned cooling laser ($\delta_0=-1.5\Gamma$) and a magnetic field of 0.5~mT applied to de-stabilise dark states. When the laser light is on (red curve), the distribution is narrower than when it is off (black curve), because Doppler cooling drives molecules towards lower transverse speeds, reducing the divergence of the beam. The authors assign a temperature that is characteristic of the transverse velocity spread, and estimate that the Doppler cooling has reduced this temperature from 50~mK to about 5~mK. The graph on the top left shows what happens when the cooling laser is blue-detuned ($\delta_0=1.5\Gamma$). In this case there is Doppler heating which increases the divergence of the beam, so increases its size at the detector. The graphs on the right show how the distributions change when the magnetic field is reduced to 0.06~mT. In this case, there is a strong accumulation of molecules at low transverse speeds when the light is blue detuned, which is characteristic of magnetically-assisted Sisyphus cooling. The authors estimate that the transverse temperature is reduced to about 300~$\mu$K. For red-detuning, there are two peaks symmetric around the centre, which correspond to molecules driven towards an equilibrium non-zero velocity. This occurs due to a competition between Doppler cooling and Sisyphus heating.

These remarkable results showed that laser cooling really could work for molecules. Soon, other groups started applying the same methods to other molecules. Transverse Doppler cooling of a beam of YO molecules was demonstrated in one and two dimensions, lowering the temperature to about 15~mK~\cite{Hummon2013}. Using the Sisyphus mechanism, a beam of SrOH was cooled to 750~$\mu$K in one transverse direction, showing that laser cooling can also work for polyatomic molecules~\cite{Kozyryev2017}. Using similar methods, a beam of YbF molecules was cooled in one dimension to a temperature below 100~$\mu$K~\cite{Lim2018}. This molecule is being used to measure the electron's electric dipole moment~\cite{Hudson2011}, and the ability to cool it to such low temperature holds great prospects for improving the precision of this measurement~\cite{Tarbutt2013}.

These transverse laser cooling experiments used molecular beams, where the initial spread of transverse velocities was just a few m/s. The experiments demonstrated that low temperatures could be reached, but only altered the velocities by this small amount. In the forward direction, the molecules were still moving at high speed, typically 100-200~m/s, with velocity spreads of around 30~m/s. The next major challenge was to use the radiation pressure of the laser light to decelerate these molecular beams to rest. This typically requires at least $10^{4}$ photons to be scattered from a counter-propagating laser beam. Only those molecules that stay inside the laser beam will decelerate, but the molecular beam is diverging, with more and more molecules moving out of the area of the laser light as time goes on. Therefore, it is important to decelerate the molecular beam as rapidly as possible, which requires a high scattering rate. This calls for high laser power, rapid de-stabilisation of dark states, and perfectly-tuned laser frequencies. The molecules are travelling towards the laser, so see a Doppler shift to higher frequency. For a typical wavelength of 600~nm and a typical speed of 150~m/s, the Doppler shift is 250~MHz, which is about 40 times larger than the typical linewidth of the laser cooling transition. This shift itself is not a problem, since the laser frequencies can be tuned to compensate. However, as the molecules decelerate, the Doppler shift changes. Typically, a velocity change of around 5~m/s changes the Doppler shift by roughly a natural linewidth ($\delta_{\rm D} \approx \Gamma$). Referring to figure \ref{LaserCoolBasic}(b), or equation (\ref{eq:Rsc}), and taking $I=I_{\rm s}$ as an example, we see that this detuning ($\delta = \Gamma$) reduces the scattering by a factor of 3. After decelerating just a small amount, the molecules are no longer resonant with the light, so stop scattering photons and stop slowing down.

\begin{figure}[t]
	\centering
	\includegraphics[width=\columnwidth]{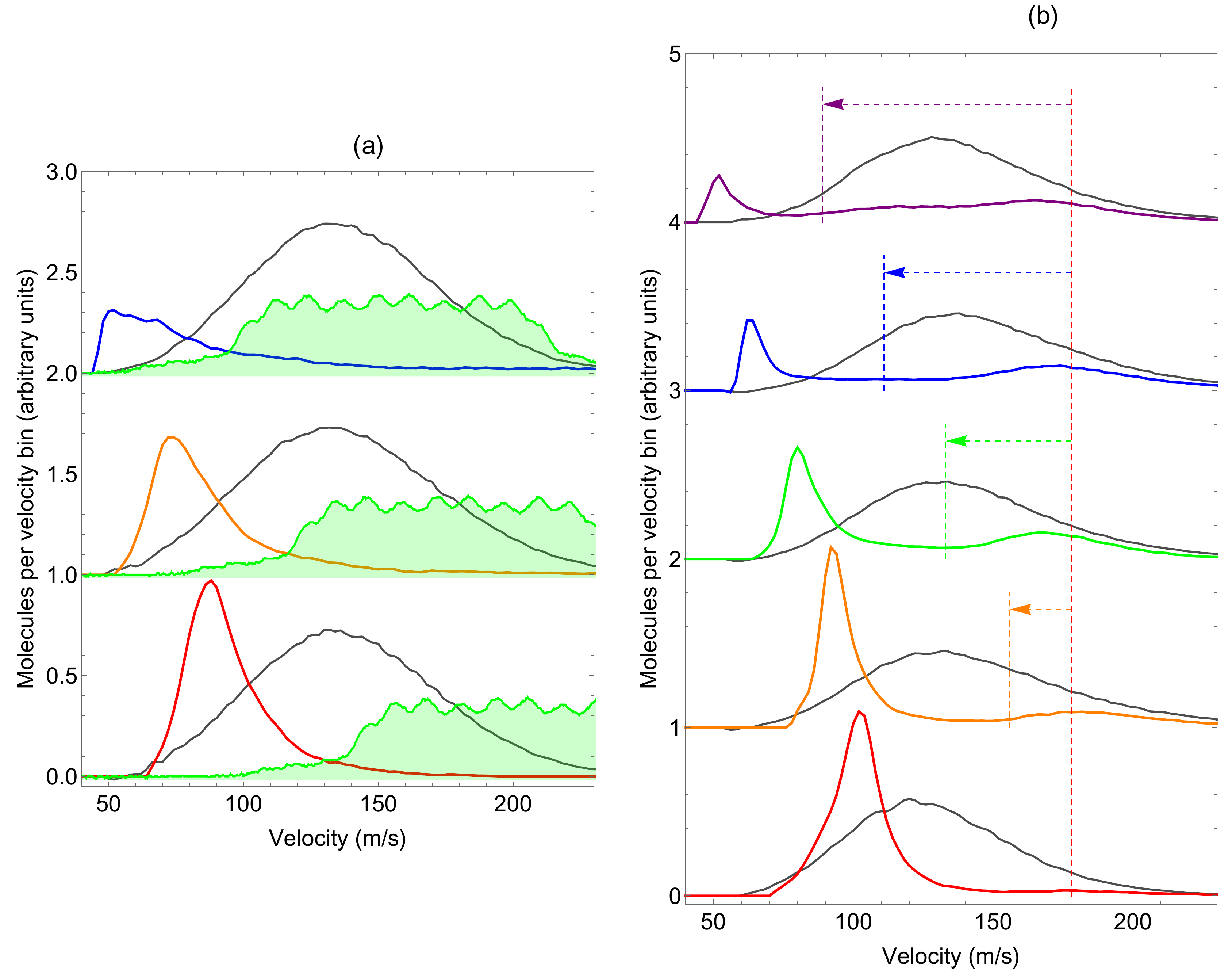}
	\caption{Slowing molecular beams using radiation pressure. Plots show velocity distributions of CaF molecules with (coloured lines) and without (grey lines) the slowing light applied. Two different slowing methods are shown. (a) Slowing using a 6~ms pulse of frequency-broadened light. The green shaded area shows the frequency spectrum of the light transposed onto the velocity scale. (b) Slowing using a 6~ms pulse of frequency-chirped light. The vertical dashed lines show the frequency of the light at the start and end of the chirp, transposed onto the velocity scale. Adapted from reference \cite{Truppe2017}.} \label{slowingData}
\end{figure}

One way to solve this problem is to broaden the frequency spectrum of the laser itself. The lasers used for laser cooling are fantastically monochromatic. The light usually has a frequency spread below 1~MHz, meaning that the frequency is defined with a precision better than 1 part per billion. This is usually desirable, because it puts all the power at the frequency that matters. However, for slowing the molecules, it helps to broaden the frequency spread of the laser to about 250~MHz so that the molecules can slow to rest without falling out of resonance. This method is sometimes called frequency-broadened slowing, and sometimes white light slowing\footnote{It's called white-light slowing because the frequency spread of the laser light is so much broader than normal. The spread is still tiny though; the light is not at all white!}. Having demonstrated the first laser cooling of molecules, the team at Yale set about slowing down their SrF beam using the white light slowing method. They used an electro-optic modulator to modulate the frequency of the laser, effectively broadening its spectrum. Starting from a beam with a mean speed of 140~m/s, they were able to slow about 6\% of the molecules to speeds below 50~m/s~\cite{Barry2012}.

Instead of broadening the frequency spectrum of the light, the frequency can be changed over time in order to compensate for the changing Doppler shift as the molecules slow down. This method, called frequency-chirped slowing, was used at Imperial College London to decelerate beams of CaF molecules~\cite{Zhelyazkova2014, Truppe2017}. The frequency-chirped method was also compared to the frequency-broadened method. Figure \ref{slowingData} shows the results of these experiments. The plots show velocity distributions of the CaF beam with and without the slowing light applied. In figure \ref{slowingData}(a) the frequency-broadened method was used. The spectrum of the laser light was broadened to about 220~MHz, corresponding to a velocity spread of about 120~m/s. Three datasets are shown. In the bottom one, the central frequency of the light is resonant with molecules moving at about 200~m/s. The molecules decelerate, accumulating at around 90~m/s. In the middle and top plots, the light is shifted to lower frequencies, causing the molecules to decelerate to lower velocities. Figure \ref{slowingData}(b) shows a sequence of experiments using the frequency-chirped method. In the bottom plot, there is no chirp, and the fixed frequency of the light is resonant with molecules moving at 178~m/s. They are decelerated by the light and accumulate at around 100~m/s. The subsequent plots show what happens as the frequency is swept linearly during the 6~ms period when the light is on. The span of the sweep is shown by the arrows in each case. As the light is swept to lower frequencies, the molecules are pushed to lower velocities. The decelerated velocity distributions are narrow, showing that the molecules have been cooled as well as slowed. The number of molecules diminishes as they slow to lower speeds, because the transverse velocity spread is unchanged, so the slower beams diverge more and fewer molecules reach the area of the detector. This highlights the importance of decelerating the beam as rapidly as possible. After optimising the frequency-chirped method, the group at Imperial were able to decelerate about $10^{6}$ CaF molecules to 15~m/s. Similar methods have been demonstrated at JILA, using YO molecules~\cite{Yeo2015}, and at Harvard with CaF molecules~\cite{Hemmerling2016}. Now that laser cooling to low temperature and laser slowing to low speed had been demonstrated, attention turned to the problem of trapping the molecules.

\section{Magneto-optical trapping of molecules}

An optical molasses provides a velocity-dependent force which damps the velocity of the atoms towards zero. On its own, this does not make a trap, and the atoms or molecules will slowly diffuse away. To make a magneto-optical trap (MOT), we add a magnetic field gradient to the optical molasses which introduces a position-dependent force, as explained in more detail below. In this way, atoms are captured, trapped for long periods, and cooled to low temperatures. Over the last 30 years, the MOT has proven spectacularly useful. It has been the starting point for almost all the innumerable experiments with ultracold atoms. It is likely to play the same role for ultracold molecules. We first show how the MOT works for atoms, then consider some problems in extending the method to molecules, and finally show how these problems have been solved.

\subsection{Principles of the magneto-optical trap}

\begin{figure}[t]
	\centering
	\includegraphics[width=\columnwidth]{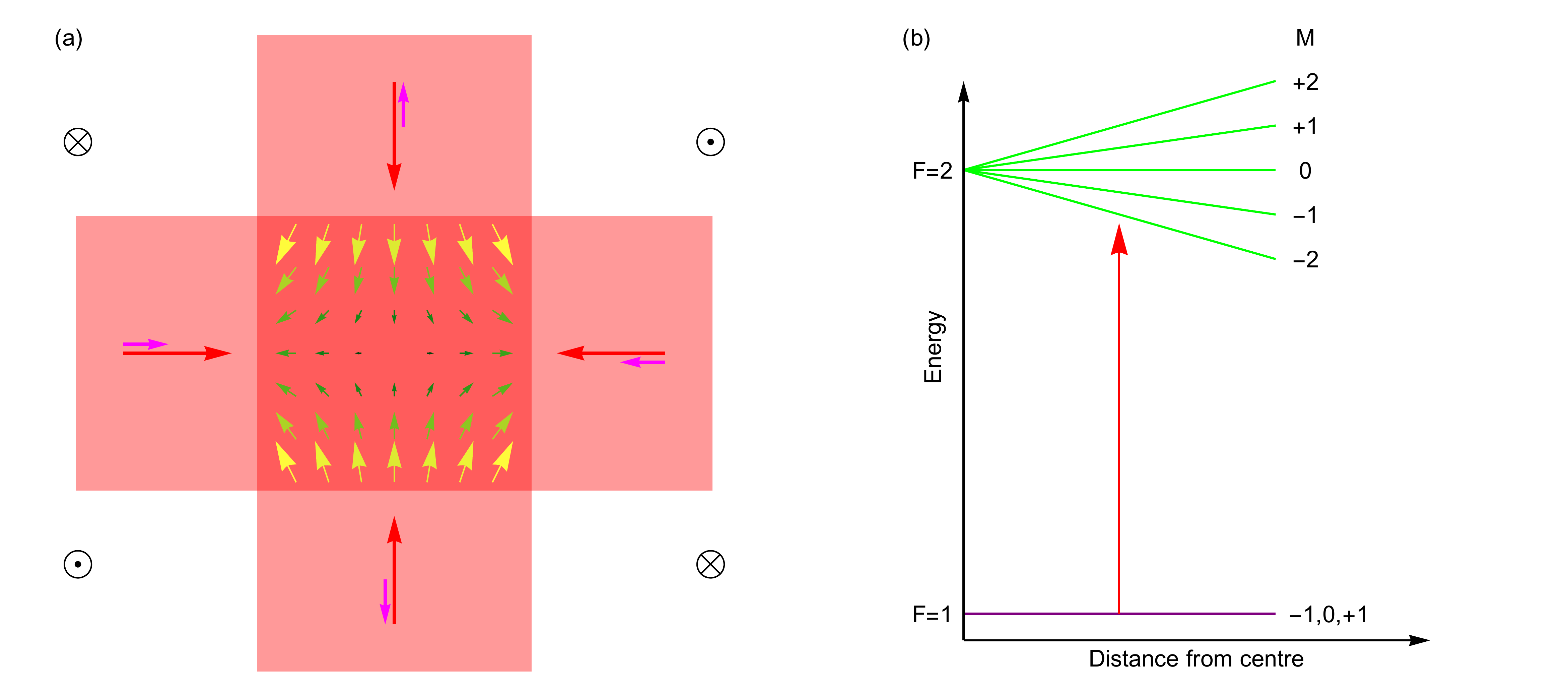}
	\caption{Magneto-optical trapping. (a) Pairs of counter-propagating, circularly-polarised red-detuned beams propagate along all three coordinate axes. Two pairs are shown by the pink regions. The long red arrows show their $k$-vectors, and the short magenta arrows indicate the directions of the photon-spin for each beam. A pair of coils, with their axes vertical, carry equal and opposite currents, producing the magnetic field indicated by coloured arrows in the region where the beams overlap. The third axis, out of the plane, is identical to the horizontal axis in the plane. In the convention used by most physicists, the laser beams in the horizontal plane are left-circularly polarised, while the vertical beams are right circularly polarised. (b) Energy levels of a model atom as a function of displacement, $z$, along any of the $k$-vectors. Whatever the displacement, the $z$-axis is along the local magnetic field direction. The model atom has a ground state with $F=1$ and no Zeeman splitting, while the $F=2$ excited state has a Zeeman splitting which increases with displacement from the centre. The red arrow indicates the energy of the photons. Those that carry angular momentum along $-z$ drive transitions that are closer to resonance, so are more likely to be absorbed.} \label{MOTBasic}
\end{figure}

Figure \ref{MOTBasic} illustrates the principle of the MOT. Pairs of circularly-polarised, red-detuned laser beams counter-propagate along each of the coordinate axes, making an optical molasses. A magnetic field is produced by equal and opposite currents flowing in a pair of coils which, in this picture, enclose the vertical beams. The strength and direction of this field are indicated by the arrows within the region where the beams overlap. The field is zero at the centre, increases linearly in all directions away from the centre, points towards the centre in the vertical direction, and points away from the centre in the horizontal direction. The propagation directions of the beams are indicated by the long red arrows, and the directions of the photon spin in each beam by the short magenta arrows. 

Consider an atom somewhere in the region where the beams intersect. In this example, we suppose that the atom has a ground state with angular momentum $F=1$ and an excited state with $F=2$, and that the excited state has a magnetic moment while the ground state does not.\footnote{The principle of the MOT is the same for all cases where $F'=F+1$, where $F$ and $F'$ are the angular momenta of the ground and excited states respectively. The principle is also unchanged if the ground state has a magnetic moment. The system in figure \ref{MOTBasic}(b) was chosen for comparison with figure \ref{MOTProblems}(a), where nothing changes other than the Zeeman splittings.} Suppose the atom is displaced from the centre, horizontally to the right. Here, the magnetic field points to the right, and we can choose this direction to be the $z$-axis. As $z$ increases, so does the magnetic field, and so too does the Zeeman splitting of the excited state. A photon coming from the right has its spin along $-z$, so absorption of this photon must reduce $M$ by 1 in order to conserve angular momentum. A photon coming from the left has its spin along $+z$, so $M$ increases by 1 when this photon is absorbed. Because of the Zeeman shift, and the negative detuning of the light, transitions that reduce $M$ are closer to resonance, so the atom is more likely to scatter photons from the right. There is a net radiation pressure which pushes the atom back towards the centre. If the atom is instead displaced along one of the other beam directions, we can again choose the $z$-axis to be in the direction of the local magnetic field, and we see that our description is unchanged. Because of the Zeeman splitting, atoms preferentially absorb photons whose spin is opposite to the magnetic field direction, and these photons always come from the beam that pushes the atom towards the centre, where the field is zero.

\subsection{Difficulties in making MOTs for molecules}

Doppler cooling in the MOT is the same as in free space. The main requirement is a closed cycling transition, which though challenging, can be engineered for some molecules as discussed in section \ref{Sec:LaserCoolingMolecules}. There are further challenges in making a MOT however. We illustrate these by describing two situations, relevant to molecules, where magneto-optical trapping fails~\cite{Tarbutt2015}.

\begin{figure}[t]
	\centering
	\includegraphics[width=0.8\columnwidth]{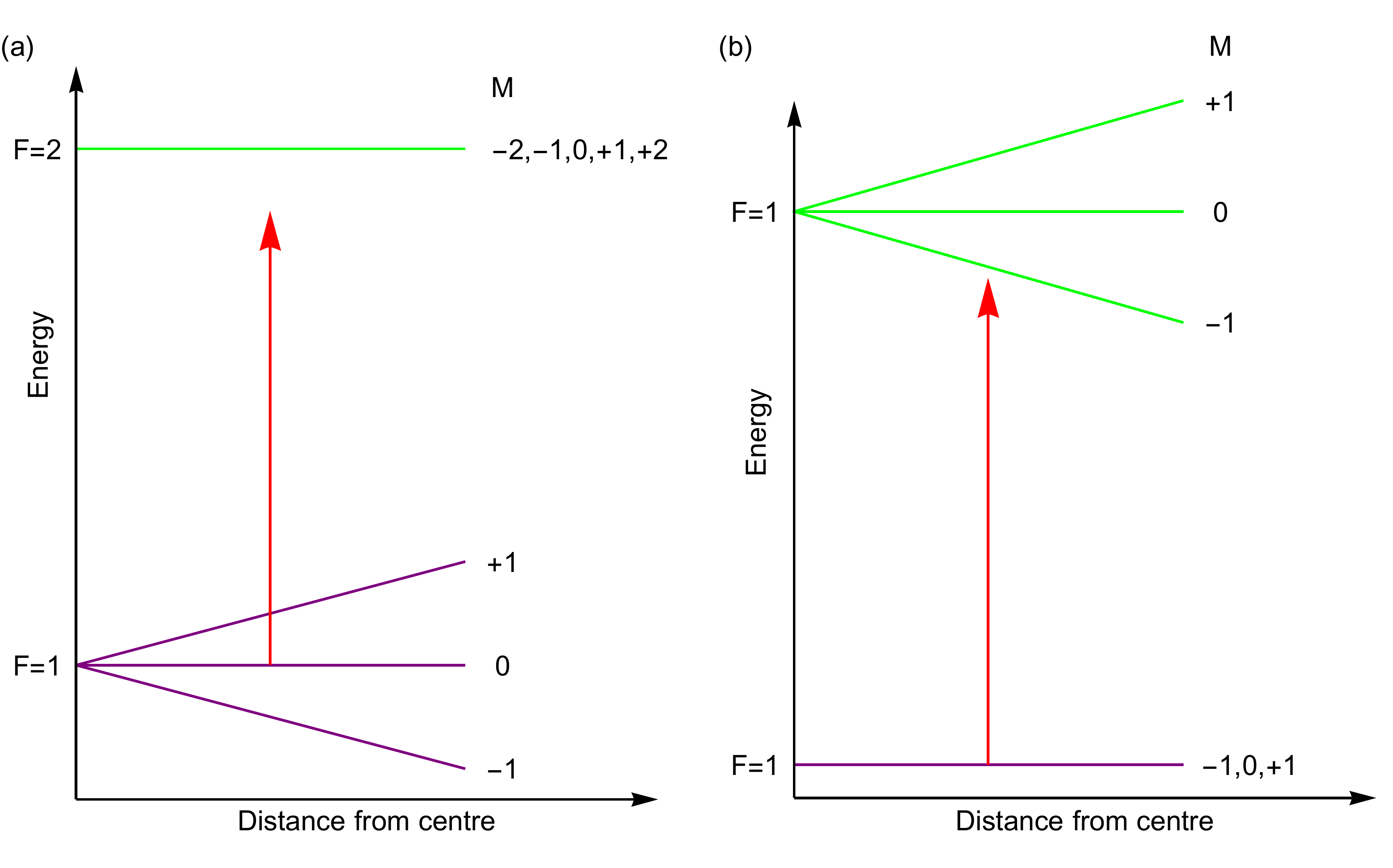}
	\caption{Configurations that do not make MOTs. (a) No Zeeman splitting in the excited state. (b) $F=1$ to $F=1$ in one dimension.} \label{MOTProblems}
\end{figure}

In figure \ref{MOTBasic}, we considered how a MOT works for an atom with a Zeeman splitting in the excited state. Now we make one small change, as illustrated in figure \ref{MOTProblems}(a): the ground state has a Zeeman splitting, but the excited state does not. As before, one laser beam drives upward transitions that reduce $M$, while the opposing beam drives transitions that increase $M$. Unlike before however, the detuning of the light is identical for these two transitions. The Zeeman splitting of the ground state does not set up any preference for absorbing photons from one beam or the other, so there can be no time-averaged confining force.\footnote{The relevant time scale is the time needed to scatter a few tens of photons. On a shorter time scale, there can be a force. For example, an atom in $M=+1$ preferentially scatters photons from the beam driving the $\Delta M=+1$ transition because it is 6 times stronger than the $\Delta M=-1$ transition. However, the atom soon reaches $M=-1$ where the transition strengths are exactly reversed. On average, the atom absorbs an equal number of photons from each beam, so there is no net momentum transfer.} The general principle is that magneto-optical trapping requires a Zeeman splitting in the excited state. Extending this idea, we find that the confining force is weak if the excited state Zeeman splitting is small compared to that of the ground state. This is precisely the situation for all the molecules where MOTs have been made so far. Consider CaF as an example. In its excited state, there are two main contributions to the magnetic moment, one due to the orbital angular momentum of the electrons, and the other due to the electron spin. These cancel almost exactly, leaving only a small residual magnetic moment. By contrast, the ground state has no orbital angular momentum, so has a large magnetic moment from the unpaired electron spin. As a result, the Zeeman splitting of the excited state is about 40 times smaller than in the ground state, too small, according to calculations, to make a useful MOT.

Next consider figure \ref{MOTProblems}(b), which shows an atom with an $F=1$ ground state and an $F=1$ excited state. The excited state has a Zeeman splitting, so we might expect the MOT to work. Consider what happens in one dimension, where there is only a pair of counter-propagating beams. A ground-state atom in $M=+1$ can only scatter a photon from one of the two beams, the one that reduces $M$, because the excited state has no levels with $M>1$. Conversely, a ground state atom in $M=-1$ can only scatter a photon from the other beam, the one that increases $M$. In both cases, the upper state has $M=0$ which spontaneously decays to the lower $M=\pm 1$ states with equal probability. Referring back to figure \ref{MOTBasic}(a), we see that an $M=+1$ atom is bright to the beam that pushes towards the centre, and dark to the opposing beam, whereas the opposite is true for an $M=-1$ atom. Because of the dark state, the atom is constrained to scatter just as many photons from one beam as from the other, and the net force is zero. This result is specific to the $F=1$ to $F=1$ case in one dimension. In three dimensions, the other beams play an important role and a trapping force is recovered. Moreover, even in one dimension, there can be a trapping force for other angular momentum cases where there are dark Zeeman levels. Nevertheless, the magneto-optical force is substantially weaker in cases where there are dark states compared to cases where there are not. This is a problem for molecules because the laser cooling transition always involves dark states, as discussed above.

In summary, we see that there are problems in making molecular MOTs, even for those species most amenable to laser cooling. The presence of dark states, and the small magnetic moment in the excited state, together lead to very weak trapping. Next, we look at two methods that circumvent these problems.

\subsection{Radio-frequency MOT}

\begin{figure}[t]
	\centering
	\includegraphics[width=0.8\columnwidth]{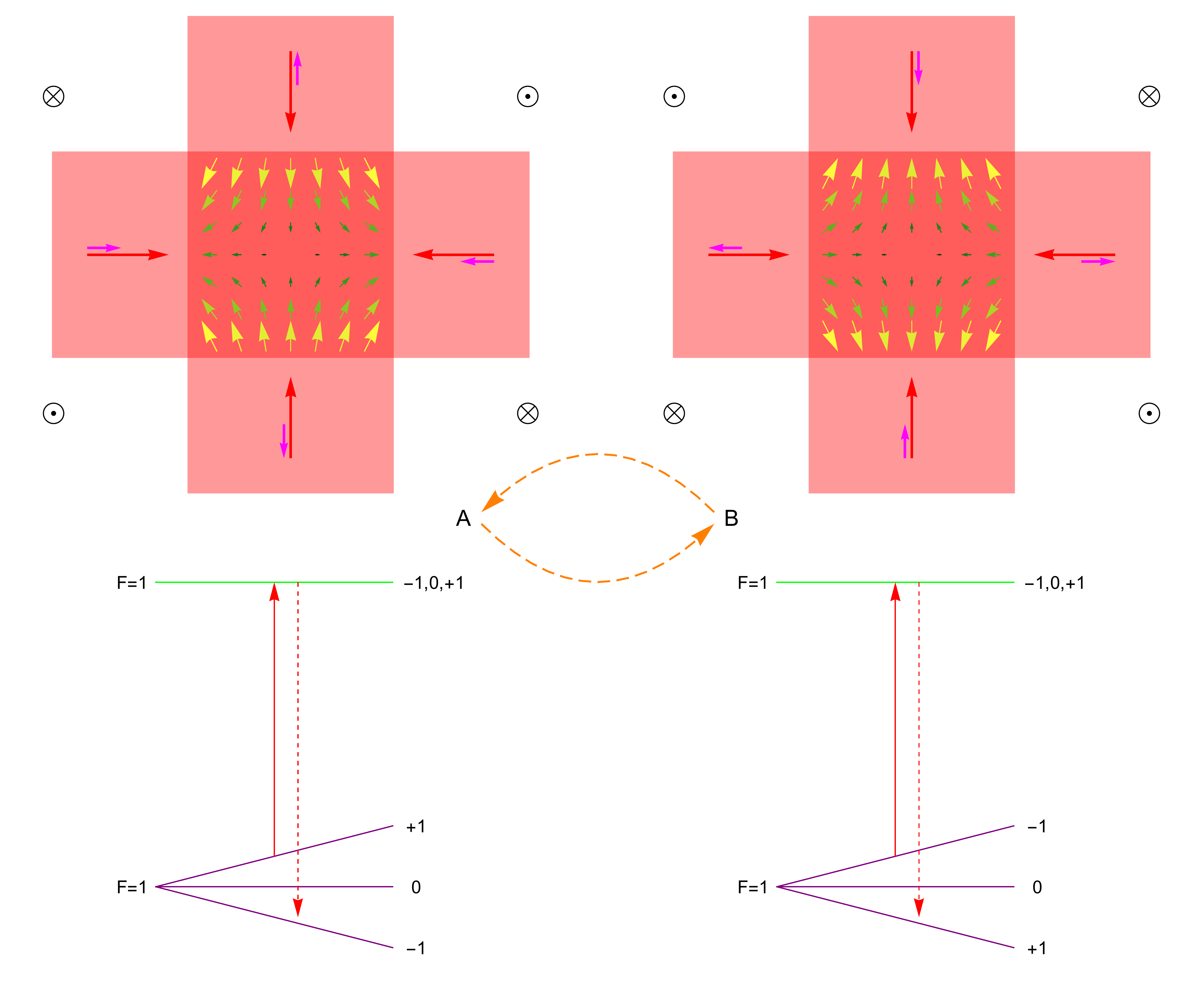}
	\caption{Principle of the radio-frequency MOT. The magnetic field and laser polarizations are switched back and forth between the two configurations shown on the left and right, labelled A and B. In A, molecules are optically pumped into $M=-1$ by the beam that pushes towards the centre. In B, they are optically pumped back to $M=+1$, again by the beam that pushes towards the centre. } \label{rfMOT}
\end{figure}

Consider a molecule with an $F=1$ ground state and an $F=1$ excited state, with no Zeeman splitting in the excited state. This is a good model for many molecules amenable to laser cooling, and is an example where a standard MOT will not work. Figure \ref{rfMOT} illustrates the principle of a radio-frequency (rf) MOT~\cite{Hummon2013, Norrgard2016}. In this scheme, the magnetic field direction and the laser polarization handedness both alternate back and forth between the two configurations shown (labelled A and B). Suppose we begin in configuration A with a molecule displaced to the right, and let us take a $z$-axis pointing to the right. Because of the beam polarizations, a molecule in $M=+1$ can scatter photons from the beam coming from the right, but not from the beam coming from the left. After scattering a few photons, it is likely to be in $M=-1$, where the situation is reversed. However, the scattering rate from $M=-1$ is slower than from $M=+1$ because the transition is further from resonance with the red-detuned light, so the molecule spends most of its time in $M=-1$. After a short time, the configuration is switched suddenly from A to B. The molecule remains in the same state, most likely $M=-1$. Because the magnetic field has reversed, so too has the Zeeman splitting, so this state is now the one closest to resonance. Since the polarizations are also reversed the molecule can, once again, scatter photons from the beam coming from the right, but not from the beam coming from the left. It soon finds itself in $M=+1$, and now the MOT is switched back to configuration A and the cycle repeats. Provided the switching is fast enough, there is strong preferential scattering from the beam that pushes the molecule back towards the centre. The switching rate should be comparable to the optical pumping rate, typically a few MHz.

\subsection{Dual-frequency MOT}

\begin{figure}[t]
	\centering
	\includegraphics[width=0.8\columnwidth]{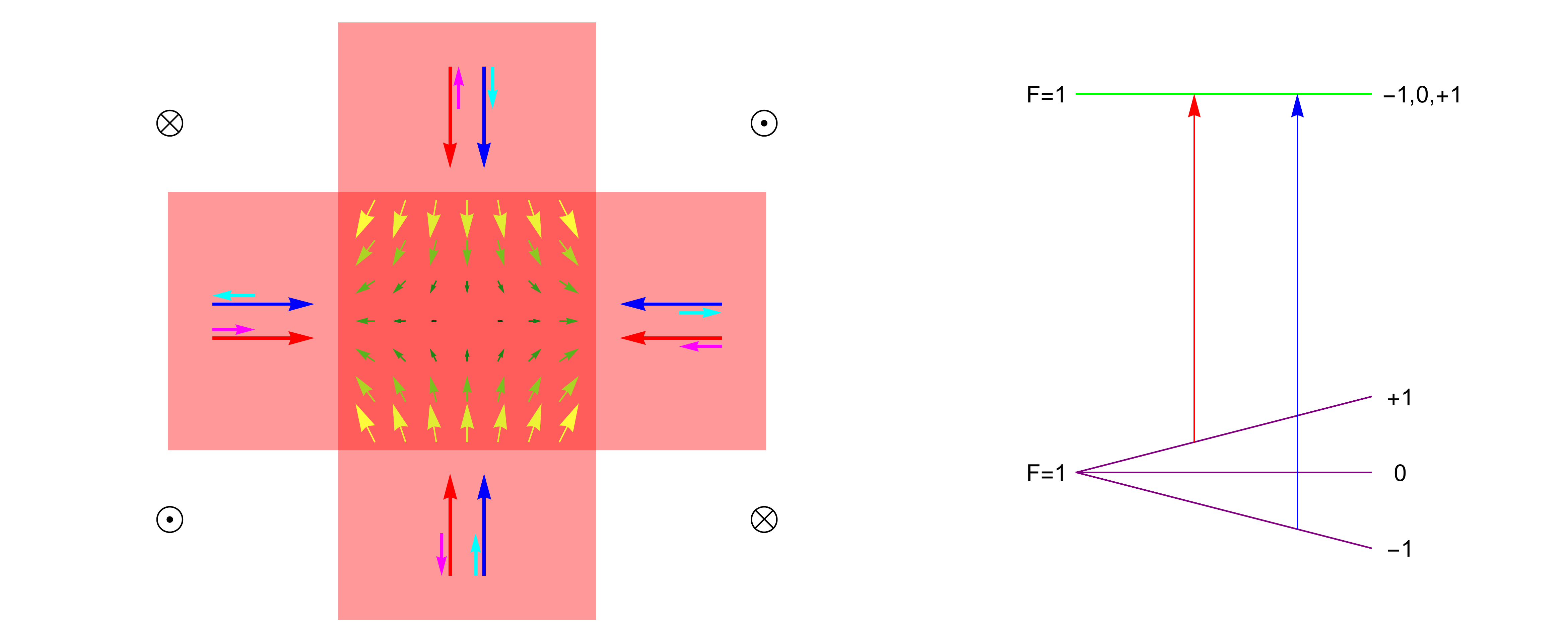}
	\caption{Principle of the dual-frequency MOT. Each beam contains two frequency components of opposite handedness, one red-detuned, the other blue-detuned. A molecule in $M=+1$ preferentially scatters photons from the red-detuned beam that pushes it towards the centre. When in $M=-1$, it preferentially scatters photons from a blue-detuned beam, again the one that pushes it towards the centre. } \label{DualMOT}
\end{figure}

Figure \ref{DualMOT} illustrates an alternative method for recovering a trapping force, known as a dual-frequency MOT~\cite{Tarbutt2015b}. Here, each of the MOT beams contains two frequency components with opposite handedness, one red detuned and the other blue detuned. Once again, consider a molecule displaced to the right, and take the $z$-axis pointing to the right, parallel to the magnetic field. A molecule in $M=+1$ preferentially scatters photons from a red-detuned component, since this is shifted closest to resonance. Of the two beams, from left and from right, the one from the right has the correct handedness to drive the transition. When the molecule is in $M=-1$ it preferentially scatters from a blue-detuned beam, and again it is the beam from the right that has the correct handedness to drive this transition. Thus, this arrangement produces a restoring force towards the centre of the MOT. However, the presence of both red- and blue-detuned light sets up a competition between Doppler cooling and heating. It turns out that Doppler cooling dominates as long as the blue-detuned component is further from resonance than the red-detuned component. In that case, there is both cooling and trapping - the MOT works.

\subsection{Progress in magneto-optical trapping of molecules}

In 2013, a group at JILA demonstrated that a beam of YO molecules could be compressed transversely by magneto-optical forces~\cite{Hummon2013}. They modulated the polarization of the cooling laser and the direction of the magnetic field gradient, implementing the radio-frequency MOT method. The group showed that, in addition to Doppler cooling of the beam to about 2~mK, the transverse distribution of molecules was compressed when the polarisation and field direction were switched in phase, but not when switched out of phase. This was an important first step towards a three-dimensional magneto-optical trap.

\begin{figure}[t]
	\centering
	\includegraphics[width=0.6\columnwidth]{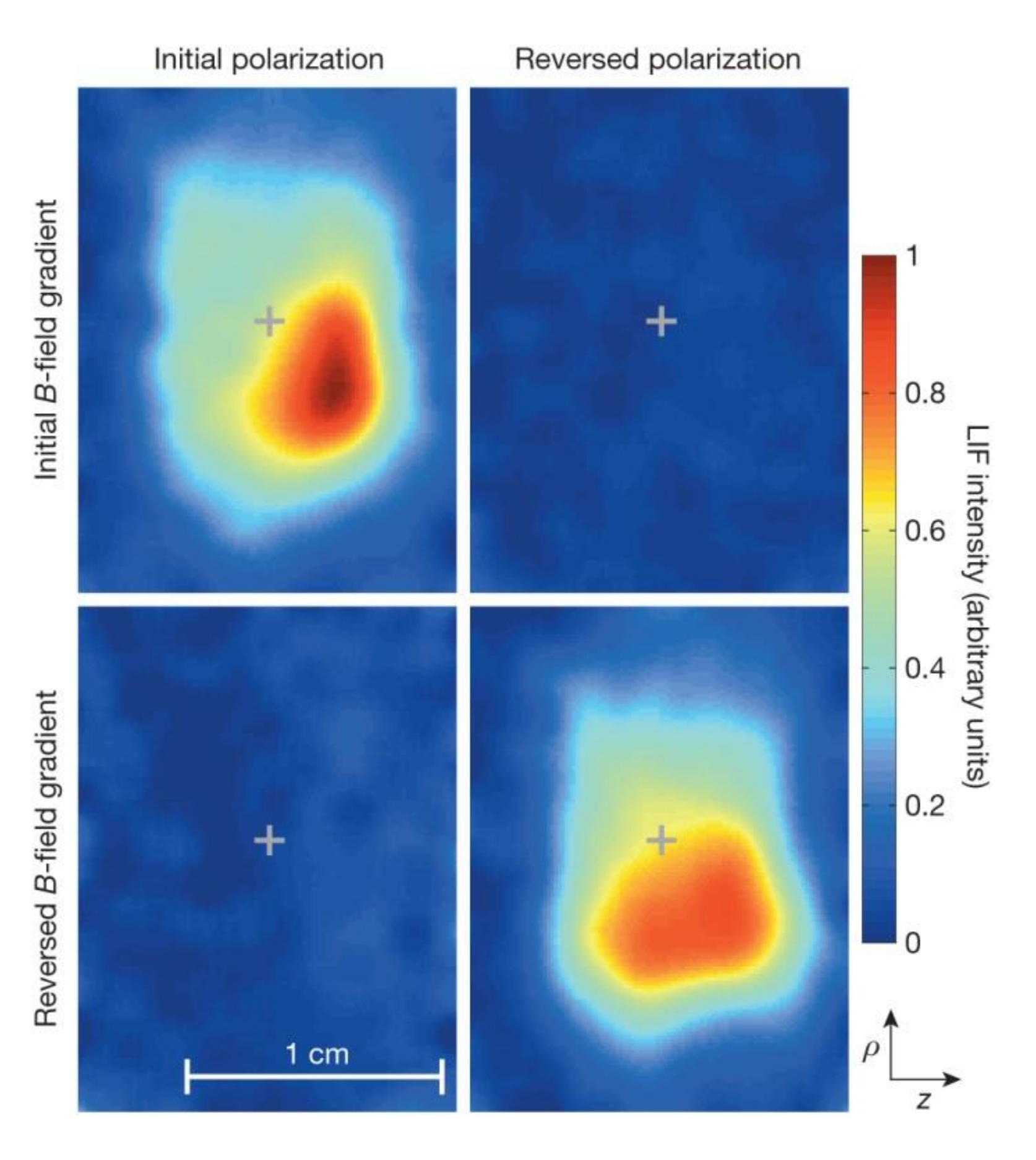}
	\caption{Magneto-optical trapping of SrF. The trapped molecules are observed by imaging their fluorescence onto a camera. The molecules are only trapped when the handedness of the laser polarisation, relative to the magnetic field direction, is chosen correctly. Reproduced from reference \cite{Barry2014} with permission from Springer Nature.} \label{SrFMOT}
\end{figure}

In 2014, the Yale group demonstrated the first 3D MOT of molecules~\cite{Barry2014}. They used SrF, following on from their earlier work on laser cooling and slowing of the same molecules. By optimising their frequency-broadened slowing technique, they produced a slow beam with a small fraction of the molecules entering the MOT region at just a few m/s, slow enough to be captured in the MOT. The cooling laser addressed the rotationally-closed component of the $X(v=0) \leftrightarrow A(v'=0)$ transition. This transition has four hyperfine components, and the frequency spectrum of the laser was carefully crafted to address each one. Three further lasers were used to close off the leaks to the $v=1$, $v=2$ and $v=3$ vibrational states. The laser-induced fluorescence from molecules captured in the MOT was imaged onto a CCD camera. Figure \ref{SrFMOT} shows images from the camera for four cases. In the top-left image, the polarizations and magnetic field direction are set to provide a magneto-optical confining force, and a bright region of fluorescence appears close to the position where the magnetic field is zero, corresponding to molecules trapped around this point. When the polarisations are reversed (top right), or the magnetic field direction reversed (bottom left), the molecules vanish because the magneto-optical forces are reversed, expelling the molecules. When both the polarisation of the light and the direction of the magnetic field are reversed, the molecules appear again. This sequence is the smoking gun of a MOT. Furthermore, the group showed that if they gave the molecules a short push, they undergo a damped oscillation around the trap centre, clearly showing that they are trapped. In these first experiments, the MOT contained about 300 molecules at a density of 600 molecules per cm$^{3}$. The temperature was 2.3~mK, the lifetime was 56~ms, and the radial trap frequency, which is a measure of the strength of the confining forces, was 17~Hz. Subsequent modelling work brought an improved understanding of the confining forces and showed how they could be increased~\cite{Tarbutt2015,Tarbutt2015b}. Implementing these ideas increased the spring constant of the trap by a factor of 20~\cite{McCarron2015}. 

These first experiments with MOTs of molecules used static polarisations and magnetic field. The next advance came in 2016, with the demonstration of the radio-frequency MOT in 3D, again using SrF~\cite{Norrgard2016, Steinecker2016}. The properties of this MOT were much better. Up to $10^{4}$ molecules were trapped, the density was increased to $2.5 \times 10^{5}$~cm$^{-3}$, the lifetime was extended to 0.5~s, and temperatures as low as 250~$\mu$K could be reached.

In 2017, the Imperial College group demonstrated a magneto-optical trap of CaF, making use of the dual-frequency method~\cite{Truppe2017b,Williams2017}. The group then showed how to cool these molecules to sub-Doppler temperatures. They transferred the molecules from the MOT into a three-dimensional, blue-detuned optical molasses where they were cooled by the Sisyphus mechanism to 55~$\mu$K~\cite{Truppe2017b}, which is well below the minimum Doppler temperature of $T_{\rm D,min} = 200$~$\mu$K. In the same year, the group at Harvard University also made a CaF MOT, using both the dual-frequency method and the radio-frequency method~\cite{Anderegg2017}. They were able to trap $10^{5}$ molecules at a density of $7\times 10^{6}$~cm$^{-3}$, the highest number and number density achieved so far. The Harvard and Imperial College groups have recently demonstrated two alternative sub-Doppler cooling schemes which cool the molecules to about 5~$\mu$K~\cite{Cheuk2018, Caldwell2018}, the lowest temperature achieved so far. Most recently, a 3D MOT of YO molecules was made at JILA~\cite{Collopy2018}, showing that a diverse range of molecular species is amenable to this method. As well as the allowed transition used to make this MOT, YO has a partly-forbidden transition with a small value of $\Gamma$, and a correspondingly small minimum Doppler temperature of 5~$\mu$K. It may be possible to cool and trap the molecules using this transition, bringing molecules to the ultracold regime directly in the MOT~\cite{Collopy2015}.

\section{Current directions}

The magneto-optical trap is a fabulous tool for accumulating, trapping and cooling molecules. However, most applications of ultracold molecules require quantum state control, which is not possible in a MOT since it relies on photon scattering, an incoherent process. Therefore, it is important to load the molecules into a conservative trap. 

Molecules prepared in a state with a positive Zeeman shift (known as a weak-field seeking state) can be trapped around a magnetic field minimum, forming a pure magnetic trap. The quadrupole magnetic field formed by the pair of coils used to make a MOT (see figure \ref{MOTBasic}) is an example where the field is zero at the centre and increases in all directions away from the centre. Recently, ultracold CaF and SrF have been stored in such pure magnetic traps~\cite{Williams2018,McCarron2018}. In both cases, after cooling to low temperature in an optical molasses, the majority of the molecules were transferred to a selected weak-field seeking state and then held in the trap for over a second. Using CaF, coherent control over the rotational state has been demonstrated by driving Rabi oscillations between the ground and first-excited rotational states~\cite{Williams2018}. Coherent superpositions of these rotational states are a key resource for applications in quantum simulation and information processing. The lifetime of these coherences has been investigated using Ramsey interferometry, both in free space and in a magnetic trap~\cite{Blackmore2018}. These first experiments demonstrated a rotational coherence time of about 1~ms and revealed a clear path to increasing this by a factor of 10 or more.

Another type of conservative trap is an optical dipole trap. This is an intense, focussed laser beam whose frequency is far below the electronic transition frequencies of the molecule. The oscillating electric field of the light induces an electric dipole moment in the molecule, oscillating in phase with the light. This induced dipole is attracted to the electric field maximum, so the molecule is pulled towards the focus of the laser beam, where the intensity is highest. If the temperature of the molecules is low enough, and the intensity of the laser is high enough, the molecules will be trapped around the intensity maximum. Because the laser frequency is tuned a long way from any molecular resonances, the photon scattering rate is very low, often well below 1 photon per second. Compared to a magnetic trap, an optical dipole trap typically has a small volume and small trap depth, but has the advantages that it is tightly confining and can trap molecules in any quantum state. This means that laser cooling can be applied within the trap, which can cool molecules towards the bottom of the trap and allow molecules to accumulate in the trap so that the density increases. At Harvard, CaF molecules have been stored in an optical dipole trap for almost a second. The molecules were cooled within the trap to about 60~$\mu$K, and the density of trapped molecules increased to $8\times 10^{7}$~cm$^{-3}$~\cite{Anderegg2018}. 

While some of the applications of ultracold molecules are already within reach, others require higher densities and lower temperatures. It is interesting to compare the de Broglie wavelength, $\lambda_{\rm dB}$ of the molecules with the average distance between them, $a$. Once $\lambda_{\rm dB}$ exceeds $a$, the molecules are in the quantum degenerate regime. At a temperature of $5$~$\mu$K, CaF molecules have $\lambda_{\rm dB}=100$~nm, while at a density of $10^{8}$~cm$^{-3}$ they have $a=22$~$\mu$m. We see that the laser-cooled samples produced so far are a long way from the quantum regime and that higher densities and lower temperatures are called for.

A good way to increase the density is to load more molecules into the magneto-optical trap. The molecule sources emit a lot of molecules, but only a tiny fraction reach the MOT with low enough speed to be captured. More efficient slowing methods are needed and several are now being developed~\cite{Fitch2016, Kozyryev2018, Wu2017, Petzold2018}. 

To reach lower temperatures, it is natural to pursue the same methods developed for ultracold atoms, especially evaporative cooling and sympathetic cooling. In evaporative cooling, the highest energy particles are removed from the thermal distribution and the remaining particles thermalise through elastic collisions to a lower temperature. Provided the collision rate is high enough, this is an efficient cooling method, and it is the standard way to produce atomic Bose-Einstein condensates. Early work on evaporative cooling of molecules has already shown promising results~\cite{Stuhl2012, Reens2017}. In sympathetic cooling, an ultracold gas acts as a refrigerant for a warmer gas of a different species, the two species thermalising through elastic collisions. Using this method, evaporatively cooled atoms could be used to cool molecules towards the quantum degenerate regime~\cite{Tokunaga2011, Lim2015}. Only the atomic gas needs to be at a high density to reach the required collision rate, and because the evaporation is applied to the atoms, all the molecules can be retained in this cooling process. If there are inelastic collisions however, molecules will be lost, so the success of this process requires a favourable ratio of elastic to inelastic collision rates. At present, there is not much experimental data on the ultracold collisions between atoms and molecules, but that will soon change with several experiments now capable of exploring this topic~\cite{Parazolli2011,Akerman2017,Williams2018,McCarron2018,Anderegg2018,Prehn2016,Wu2017}.
 
An exciting direction for the field is the exploration of many-body quantum systems using arrays of polar molecules interacting through the dipole-dipole interaction. It is ideal to start with a single molecule on each site of an array, with spacings small enough for strong dipole-dipole interactions, and the possibility of introducing a controlled degree of disorder. To make an array, one approach is to use the interference of overlapping laser beams to make an optical lattice that can trap molecules at each antinode. This is commonly done for atoms, and has been done more recently for ultracold molecules formed by association of ultracold atoms~\cite{Yan2013}. For a high filling fraction it is necessary to start from a gas that is close to quantum degeneracy. Although laser-cooled molecules are still far from this regime, the advances outlined above may soon bring them there. An alternative approach for making small arrays of molecules, that does not demand such high initial densities, is to use tightly-focussed optical dipole traps, usually called tweezer traps. A single tweezer trap is made by focussing a laser as tightly as possible so that the spot size is about the same as the wavelength of light. A single molecule from an optical molasses can be cooled into the trap following the approaches already developed for atoms~\cite{Schlosser2001,Kaufman2012, Barredo2016, Bernien2017, Liu2018}. If a second molecule enters when the molasses light is on, it is likely that the two will collide on a short timescale, ejecting both, so the trap almost always contains either zero or one molecule. In Harvard, single molecules were recently trapped this way~\cite{Anderegg2019}. An array of these traps can be made, and each trap in the array can be moved, making it reconfigurable. Imaging the fluorescence from trapped molecules onto a CCD camera determines the occupancy of each site in the array, and the occupied traps can then be moved to make the desired configuration. This approach may be more versatile than the optical lattice approach, and does not demand such high initial densities, but it is probably limited to quite small arrays.

As outlined in section \ref{Sec:Intro}, molecules can be used to make very precise tests of fundamental physics. For example, molecules are already used to make the best measurements of the electron's electric dipole moment (edm)~\cite{Hudson2011,Baron2014,Cairncross2017,Andreev2018}. These measurements are sensitive to new forces that violate time-reversal symmetry and place tight constraints on many proposed extensions of the Standard Model. The extraordinary precision of these experiments can be further improved by using heavy, laser-coolable, polar molecules, and experiments using ultracold YbF, BaF and polyatomic molecules such as YbOH, are currently being built~\cite{Lim2018, Aggarwal2018, Kozyryev2017b}. New interactions that violate time-reversal symmetry can also be probed by measuring the edm of the proton, and an experiment to do this using laser-cooled TlF molecules is underway~\cite{Norrgard2017}. Advances in laser cooling of molecules, such as the development of laser-cooled molecular fountains~\cite{Tarbutt2013, Cheng2016} are likely to have an impact on studies of parity violation in nuclei and in chiral molecules, and on searches for changes in the fundamental constants such as the fine-structure constant and the electron-to-proton mass-ratio. In these ways, laser-cooled molecules have the potential to probe new theories of physics that lie beyond the collision energy of the Large Hadron Collider or any envisaged future accelerator.

\section*{Acknowledgement(s)}
I am grateful to Luke Caldwell for helping to improve this paper.

\section*{Funding}
This work was supported by the EPSRC under grants EP/M027716/1 and EP/P01058X/1, and by a grant from the John Templeton Foundation.

\bibliography{references}

\end{document}